\documentclass[
    ,final            
  ]
  {aipproc}

\layoutstyle{6x9}

\input{psfig.sty}

\begin{document}

\title{Lectures on Dark Energy \\ and Cosmic Acceleration}


\author{Joshua A. Frieman}{
  address={Fermilab Center for Particle Astrophysics, Batavia, IL 60510 \\
Kavli Institute for Cosmological Physics, The University of Chicago, Chicago, IL 60637}
}

\begin{abstract}
The discovery ten years ago that the expansion of the Universe 
is accelerating put in place the present cosmological model, in which the Universe is composed of 4\% baryons, 20\% dark matter, and 76\% dark energy. Yet the underlying cause of cosmic acceleration remains a mystery: it could arise from the repulsive gravity of dark energy -- for example, the quantum energy of the vacuum -- or it may signal that General Relativity breaks down on cosmological scales and must be replaced. In these lectures, I 
present the observational evidence for cosmic acceleration and what it has revealed about dark energy, 
discuss a few of the theoretical ideas that have been proposed to explain acceleration, and describe the 
key observational probes that we hope will shed light on this enigma in the coming years. 
Based on five lectures given at the XII Ciclo de Cursos Especiais at the 
Observatorio Nacional, Rio de Janeiro, Brazil, 1-5 October 2007.
\end{abstract}

\maketitle


\section{Introduction}

In 1998, two teams studying distant Type Ia supernovae
presented independent evidence that the expansion of the Universe is speeding
up \citep{riess98,perlmutter99}. Subsequent observations, including more 
detailed studies of supernovae and independent evidence from the cosmic 
microwave background, large-scale structure, and clusters of galaxies, 
have confirmed and firmly established this remarkable finding. These 
lectures\footnote{The slides of the lectures are available at \url{
http://www.on.br/institucional/portuguese/dppg/cpgastron/ciclo2007/lectures.html}.}
provide a short pedagogical overview of dark energy, including: 
1. A brief review of cosmology; 2. The discovery of cosmic acceleration 
from observations of distant supernovae, followed by measurements of 
the cosmic microwave background and of baryon acoustic oscillations (BAO) in 
large-scale galaxy clustering; 3. 
The current status of the evidence for cosmic acceleration; 4. Theories  
of acceleration, including dark energy, the cosmological constant, 
and modified gravity; 5. Probes of dark energy, including clusters 
and weak gravitational lensing in addition to supernovae and BAO; and 6. A summary of 
current and future projects aimed at probing acceleration 
through the history of cosmic 
expansion and the growth of large-scale structure. 

A number of useful reviews target different aspects of the subject, including:   
theory \citep{Copeland_review,Padmanabhan_review}; cosmology \citep{Peebles_Ratra_03};  
the physics of cosmic acceleration \citep{Uzan_06}; probes of dark energy 
\citep{Hut_Tur_00}; dark energy reconstruction \citep{Sahni_review}; dynamics
of dark energy models \citep{Linder_review}; the
cosmological constant \citep{carroll92,Carroll_LivRevRel}, and the 
cosmological constant problem \citep{WeinbergRMP}. These lecture notes most 
closely follow (and borrow from) the recent review of \citet{fth08}.

\section{Brief Review of Cosmology}

The Friedmann-Robertson-Walker (FRW) cosmological model provides the
context for interpreting the observational evidence for cosmic
acceleration as well as the framework for understanding how
cosmological probes in the future will help uncover the cause of acceleration.
For further details on basic cosmology, see, e.g., the textbooks of
\citet{Dodelson_book,Kolb_Turner,Peacock_book}, and \citet{Peebles_book}.

The Universe is filled with a bath 
of thermal radiation, the cosmic microwave background radiation (CMB). 
The CMB has a purely thermal spectrum to the precision with which it has 
been measured so far, with a temperature of $T=2.725$ K above 
absolute zero. On all angular scales, it is observed to be 
nearly isotropic (rotationally invariant) 
around us, with temperature fluctuations on the 
order of $\delta T/T \sim 10^{-5}$. CMB photons have been 
travelling freely since the epoch of {\it last scattering}, 
when ionized plasma (re)combined to form neutral hydrogen, 
around 380,000 years after the Big Bang. To first approximation, 
maps of the CMB give us a picture of conditions at this early time 
and show that the young Universe was quite smooth.

According to the {\it Cosmological Principle}, also called the 
Copernican Principle, we are not priviledged observers. Therefore, 
the Universe should appear quasi-isotropic, when averaged over 
large scales, to all similar observers. (Here, similar observers 
are those moving slowly with respect to the comoving coordinates 
introduced below.) A Universe that appears isotropic to all such 
observers can be shown to be {\it homogeneous}, that is, 
essentially the same at every location, 
again averaged over large scales. More specifically, the density 
of the cosmic fluid, when averaged over scales larger than 
$\sim 100$ Mpc, is approximately translation-invariant.
 
\subsection{The Expanding Universe}

The only time-dependent degree of freedom 
which preserves homogeneity and isotropy is overall 
expansion or contraction. Preservation of this high degree of symmetry 
implies that the Universe on the largest scales is described by a single 
degree of freedom, the cosmic scale factor, $a(t)$, where $t$ is 
cosmic time. The spatial hypersurfaces at fixed time can be described 
in terms of expanding or comoving coordinates, much like the points of 
constant longitude and latitude on an expanding spherical balloon. 
To first approximation, i.e., neglecting small peculiar velocities, 
galaxies are at rest in these comoving coordinates, and the scale 
factor describes the time dependence of their physical separations, 
$d=a(t) r$, where $r$ is the fixed comoving distance between them. 
Comparing the physical separations at times $t_1$ and $t_2$, the 
apparent recession speed is given by

\begin{equation}
v = {{d(t_2) - d(t_1)} \over {t_2 - t_1}} = {{r[a(t_2)-a(t_1)]} \over
{t_2 - t_1}} = {d \over a}{da \over dt} \equiv d H(t) \simeq dH_0 ~,
\label{eq:vrec}
\end{equation}

\noindent where $H(t)\equiv (1/a)(da/dt)=\dot{a}/a$ is the  
expansion rate, $H_0$ is the Hubble parameter, the 
present value of $H(t)$, and 
the final equality in Eqn.\ref{eq:vrec} holds for small time intervals, $t_2 - t_1 \ll 1/H_0$. 
We use the subscript '0' on a quantity to denote its value at the 
present epoch.

The physical wavelengths of radiation scale with the scale 
factor, $\lambda \sim a(t)$. As a result, in an expanding  
Universe, 
light emitted by one observer at time $t_1$ and observed by 
another at a later time $t_2$ is observed to be shifted to 
longer, redder wavelengths. The redshift $z$ is thus defined by 
 
\begin{equation}
1+z = {\lambda(t_2)\over \lambda(t_1)} = {a(t_2)\over a(t_1)} ~,
\label{eq:redshift}
\end{equation}

\noindent and directly yields the relative size of the Universe at the 
time of emission. For nearby galaxies, the redshift is related to 
the apparent recession velocity by $z \simeq v/c$. 
In optical surveys, more distant galaxies  
have their light shifted farther to the red:  
they emitted their light when the Universe was smaller, indicating 
that it has been expanding. 

In the early 1920's, Slipher reported spectroscopic measurements of 
recession velocities for 40 relatively nearby spiral galaxies. In the late 
1920's, Hubble found Cepheid variable stars in $\sim 20$ nearby 
galaxies and measured their periods of variability. Using the 
period-luminosity correlation previously found by Henrietta Leavitt for 
Galactic Cepheids, he was able to infer their luminosities. From 
measurements of their apparent brightnesses, he could thus deduce 
their distances. Comparing these distances with Slipher's radial 
velocities, Hubble found empirically that $v=H_0 d$, in agreement 
with the prediction for an expanding universe (Eqn. \ref{eq:vrec}). 
While Hubble's measurements were confined to the relatively 
local universe, $d \sim $few Mpc, where peculiar velocities 
due to large-scale structure, of order 
300 km/sec, are comparable to the expansion velocity, 
modern observations have extended the measurement to much larger 
distances. The  
Hubble Space Telescope Key Project used measurements 
to $d \sim 300$ Mpc and found $H_0 = 72 \pm 8$ km/s/Mpc 
\citep{Freedmanetal}.

\subsection{Expansion Dynamics}

How does the expansion of the Universe change over time? Since gravity 
dominates over other forces on the largest scales, one expects that 
mutual gravitational attraction of the matter in the Universe would 
lead to a slowing of the expansion over time. We can put more meat 
(preferably churrasco) on this statement by considering cosmological 
dynamics. 

A Newtonian treatment captures the essence of the argument. Consider a test 
mass $m$ a distance $d$ from the center of a homogeneous, spherical ball of 
matter that we imagine carving out from the Universe. The total mass 
of the ball is $M=(4\pi/3)\rho d^3$, where $\rho$ is the density of 
the Universe. By Newton's theorem, the total energy of the test mass 
is given by 

\begin{equation}
E= {m v^2 \over 2} - {G M m \over d}~,
\label{eq:Newt}
\end{equation}

\noindent and is conserved. Using Hubble's law, $v=Hd$, we find the 
Friedmann equation,

\begin{equation} 
{2E \over md^2} \equiv -{K\over a^2(t)} = H^2(t) - \left({8 \pi \over 3}
\right) G \rho(t) ~,
\label{eq:Fried}
\end{equation}

\noindent where $K$ is a constant. In General Relativity, $K$ emerges 
as the indicator of the global spatial curvature of constant-time 
hypersurfaces: $K=0$ for flat, Euclidean space, $K>0$ for positively 
curved, spherical 
geometry, and $K<0$ for a negatively curved, three-dimensional 
hyperboloid or saddle. 

The full General Relativistic Friedmann equations for a 
multi-component Universe can be written

\begin{equation}
H^2(t) \equiv \left({1\over a}{da \over dt}\right)^2 = {8 \pi G \over 
3} \sum_i \rho_i(t) - {k \over a^2(t)}~,
\label{eq:F1}
\end{equation}

\noindent where $k=0,+1,-1$ for zero, positive, and negative curvature, and 

\begin{equation}
{1\over a} {d^2a\over dt^2} = -{4\pi G \over 3}\sum_i \left(\rho_i + 3p_i\right)~,
\label{eq:F2}
\end{equation}

\noindent where $p_i$ and $\rho_i$ are the pressure and density 
of the $i$th component. From 
Eqn. \ref{eq:F1}, it is convenient to define the critical density for 
a spatially flat Universe, $\rho_{crit} \equiv {3H_0^2/8\pi G} = 
1.88 h^2 \times 10^{-29}$ gm/cm$^3$, where the dimensionless Hubble 
parameter $h=H_0/(100 ~{\rm km/sec/Mpc})$. The present density parameter 
in the $i$th component is then $\Omega_{i,0}=\rho_i(t_0)/\rho_{crit}$.

The Einstein 
equations also imply conservation of stress-energy, which takes the 
form

\begin{equation}
{d\rho_i\over dt}+ 3\left(\rho_i + p_i\right)H=0~.
\label{eq:cons}
\end{equation} 

\noindent If the pressure and energy density are related 
by $p_i=w_i \rho_i$, 
then the density in the $i$th component scales as 

\begin{equation}
\rho_i \propto \exp \left[3 \int_0^z [1+w_i(z')]d \ln (1+z')\right]~.
\label{eq:rhow}
\end{equation}

\noindent If the equation of state parameter 
$w_i$ is time-independent, then 

\begin{equation}
\rho_i \sim a^{-3(1+w_i)}~.
\label{eq:aw}
\end{equation}
 
\noindent For non-relativistic matter, $w_m \sim (v_m/c)^2 \ll 1$, and 
$\rho_m \sim a^{-3}$; for radiation or more 
generally for ultra-relativistic particles, $w_r=1/3$, and 
$\rho_r \sim a^{-4}$. Currently, $\Omega_m \simeq 0.25$ and 
$\Omega_r \simeq 0.8 \times 10^{-4}$, implying that the Universe was 
radiation-dominated at epochs for which $1+z> a(t_0)/a(t_{eq}) 
\simeq 3000$. 

For much of the 1980's and 90's, a burning question was whether 
the Universe would expand forever or recollapse. For a 
matter-dominated universe, $w_{tot}=w_m \simeq 0$, 
``geometry is destiny'': from Eq. \ref{eq:F1}, 
positive curvature, i.e., $k>0$, implies 
$\Omega_{m,0}>1$, and since $\rho_m$ drops more rapidly than 
the spatial curvature term as $a(t)$ becomes large, the Universe 
reaches a maximum size ($H=0$) and subsequently 
recollapses. If $k \leq 0$, then 
$\Omega_{m,0} \leq 1$, and $H(t)$ remains positive: 
the Universe expands forever. 
Moreover, from Eqn. \ref{eq:F2} we have $\ddot{a} < 0$ in 
all cases: the expansion of the 
Universe decelerates due to gravity, as one expects. 

\subsection{Cosmic Acceleration and Dark Energy}

In 1998, two groups observing distant supernovae found evidence that 
the expansion of the Universe is instead speeding up, $\ddot{a}>0$ \citep{riess98,perlmutter99}.  
Since then, evidence has accumulated that the Universe was 
decelerating at early times but began accelerating about five 
billion years ago. Logically, there are three possible modes 
of explanation for this behavior: (i) we posit a form 
of ``gravitationally repulsive'' 
stress-energy in the Universe, now called Dark Energy, 
which came to dominate over 
non-relativistic matter about 5 billion years ago; 
(ii) we instead modify the geometric as 
opposed to the stress-tensor components of the Einstein-Hilbert action of  
General Relativity, a modification primarily manifest only on 
cosmologically large scales; (iii) we leave General Relativity 
and the matter-dominated Universe intact and instead drop the 
assumption that the Universe is spatially homogeneous on large 
scales, invoking large-scale structure to induce an apparent 
acceleration. Either of the first two would have profound 
implications for our understanding of fundamental physics. 
We will return to some of these theoretical ideas in a later 
chapter.

\begin{figure}
\centerline{\hspace*{0.4cm}
\psfig{file=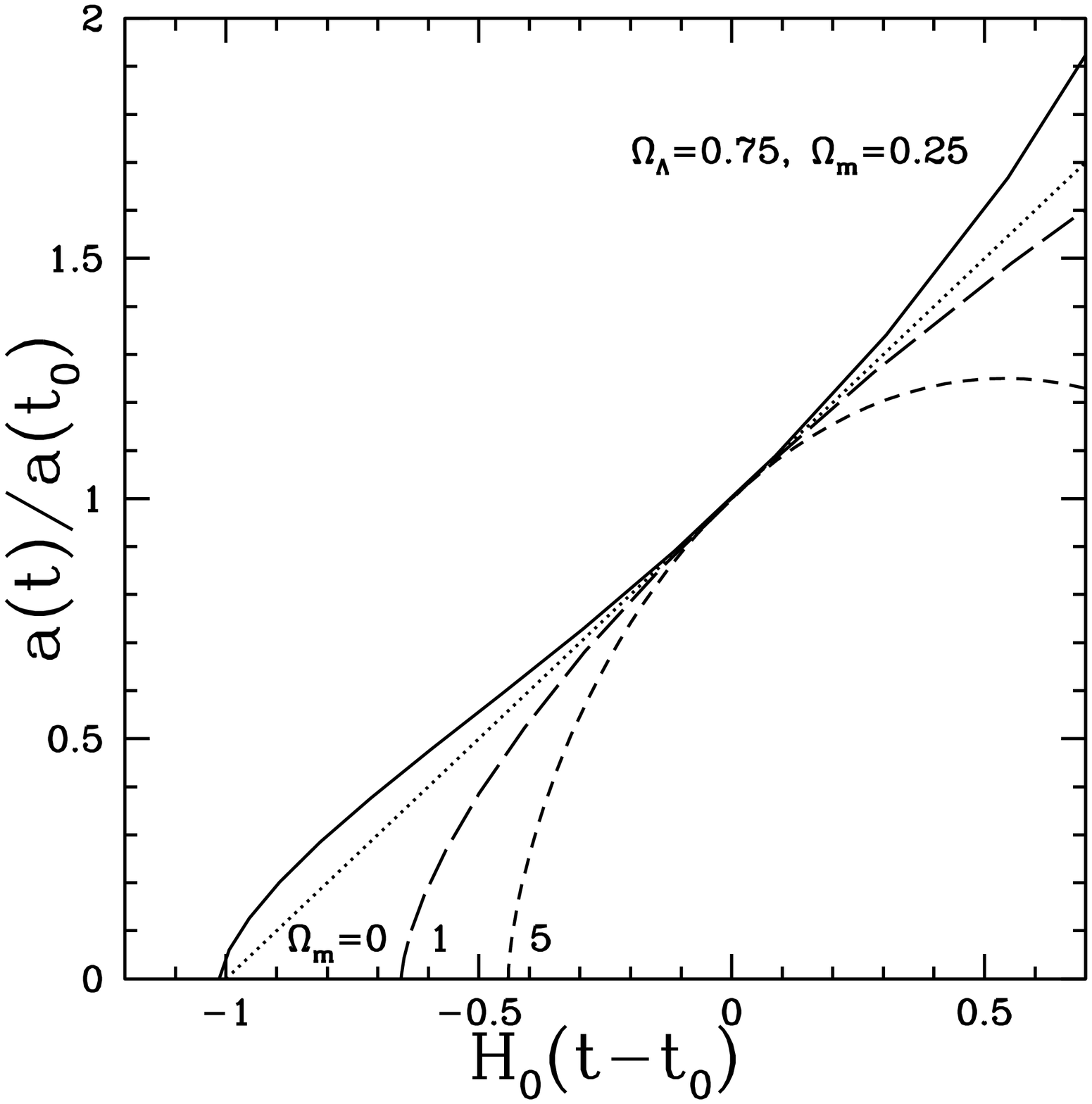,height=2.7in}
\hspace*{-0.25cm}
\psfig{file=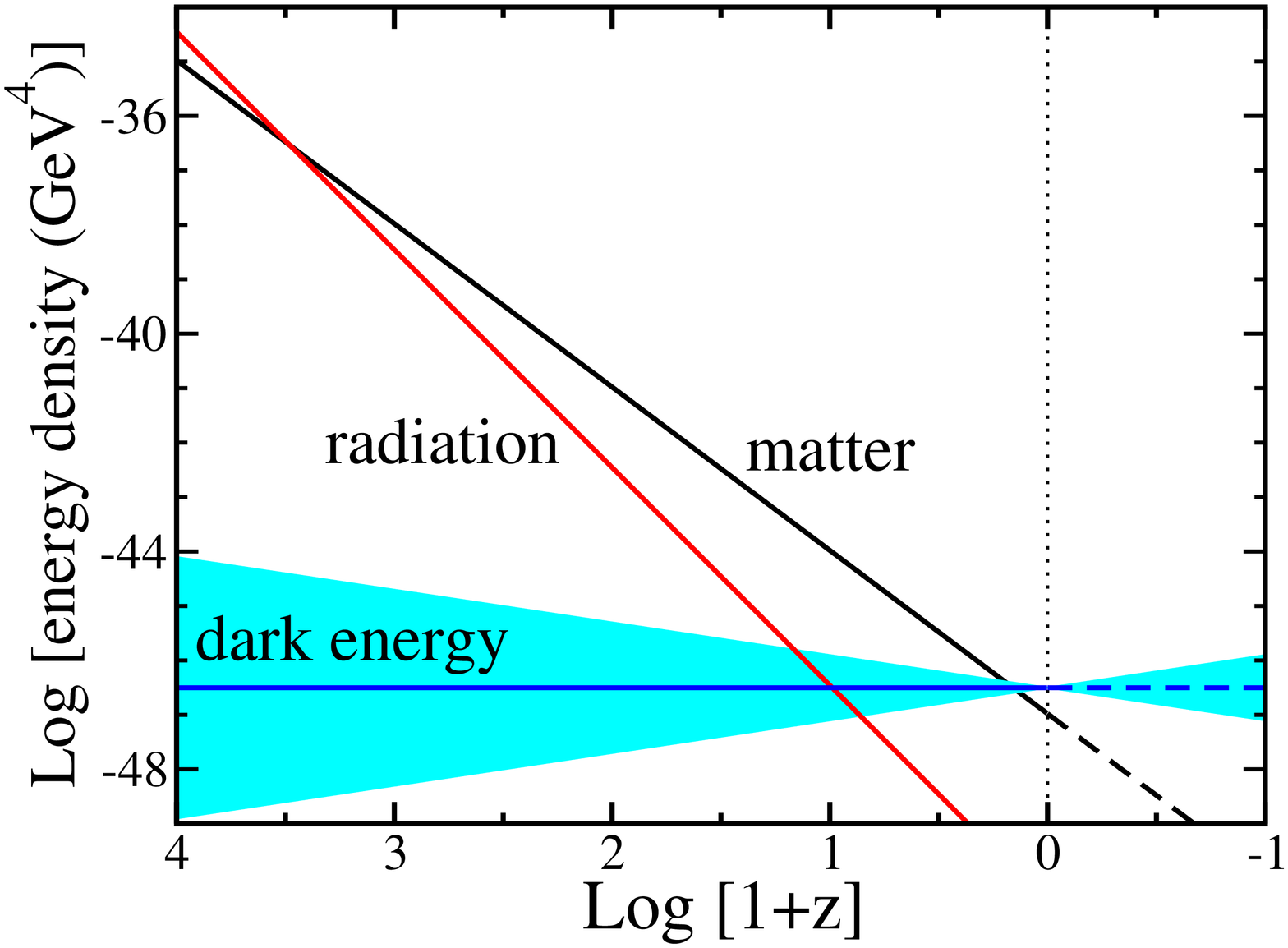,height=2.7in,angle=-90}}
  \caption{Left panel (a): Evolution of the scale factor vs. time 
for four cosmological models: three matter-dominated models with 
$\Omega_0=\Omega_m=0, 1, 5$, and one with $\Omega_\Lambda=0.75, \Omega_m=0.25$.
Right panel (b): Evolution of radiation, matter, and dark energy densities with 
redshift. For dark energy, the band represents $w=-1 \pm 0.2$. From 
\citet{fth08}.}
\label{fig:scale}
\end{figure}

Dark energy is perhaps the simplest explanation for cosmic 
acceleration and the most familiar. From Eqs. \ref{eq:F2} and 
\ref{eq:aw}, 
if a component has an equation of state parameter 
$w<-1/3$, i.e., sufficiently negative pressure, then it will 
come to dominate over other forms of stress-energy and 
will drive accelerated expansion. This is the defining 
property of dark energy. An immediate consequence is that 
the link between geometry and destiny is broken: for example, while 
$\Omega_0>1$ still implies positive spatial curvature, $k>0$, 
it does not mean that the Universe will necessarily 
recollapse, because the dark energy density scales more 
slowly with $a(t)$ than the spatial curvature term, which goes 
as $1/a^2(t)$. The FRW scale factor vs. time is shown in Fig. 
\ref{fig:scale}a for various cosmological parameter choices.
The history of the matter, radiation, and 
dark energy components is shown in Fig. \ref{fig:scale}b.

The simplest candidate for dark energy is the cosmological 
constant, $\Lambda$. It was introduced by Einstein into 
the equations of General Relativity in order to produce 
a static, finite Universe: 

\begin{equation}
G_{\mu\nu} - \Lambda g_{\mu\nu} = {8 \pi G \over c^4} T_{\mu\nu}
\label{eq:Ein}
\end{equation}

\noindent where $G_{\mu\nu}$ is the Einstein tensor describing 
the curvature of spacetime, and $T_{\mu\nu}$ is the stress-energy 
tensor of the components. When 
the expansion of the Universe was discovered by Hubble, 
Einstein advocated abandoning the cosmological constant as 
an unnecessary blemish on his theory. However, in the late 1960's, 
Zel'dovich stressed that $\Lambda$ logically belongs on the 
right-hand side of the Einstein equations, as the stress-energy 
of the vacuum, $T_{\mu\nu}^{vac} = (\Lambda/8\pi G)g_{\mu\nu}$ \citep{zeldovich68}. 
(This point was made earlier by Lemaitre as well.) Since the 
vacuum energy density gets contributions from quantum 
zero-point fluctuations of all fields, it cannot simply 
be dismissed. The pressure 
and energy density of the vacuum can be read off from the 
stress-tensor as 
$p_{vac} = - \rho_{vac} = \Lambda/8 \pi G$, so it acts 
as a fluid with equation of state parameter $w=-1$, as needed 
to explain acceleration. In a $\Lambda$-dominated model, 
the expansion is asymptotically exponential, $a(t) \sim \exp(\sqrt{\Lambda/3} t)$.
The one fly in the ointment is that 
the required energy density for cosmic acceleration is 
of order $\rho_{vac} \simeq (0.003 ~{\rm eV})^4$, while 
estimates of the vacuum energy density of quantum fields 
are at least 60 to 120 orders of magnitude larger. This 
embarrassing discrepancy, which predates and is logically 
separate from but is brought into focus by cosmic acceleration, is 
known as the cosmological constant problem.

While the discovery of cosmic acceleration is often portrayed 
as a surprise, in fact it fit neatly into a pre-existing 
theoretical and observational framework that had been 
solidifying throughout the 1990's. There were several elements 
of this framework: (1) the theory of primordial inflation \citep{guth80}
predicted a flat Universe, $\Omega_0=1$, while observations 
of dark matter 
were pointing with increasing accuracy to $\Omega_m \simeq 0.25$, 
so a component of ``missing energy'' with $\Omega_{ME} \simeq 
0.75$ was needed to reconcile the two; (2) such a missing 
energy component must be smoothly distributed and would 
therefore inhibit the growth of large-scale structure---it must 
therefore have come to dominate over non-relativistic matter 
at recent cosmic epochs, which means it must have a sufficiently 
negative equation of state parameter, $w \leq -0.5$; 
(3) the model of structure formation with cold dark matter 
and a cosmological constant, $\Lambda$CDM, in combination with 
primordial perturbations from inflation, had been found 
to be in good agreement with observations of the large-scale 
clustering of galaxies, e.g., as observed in the APM survey \citep{efstathiou90};  
(4) estimates of globular cluster ages, in combination with 
Hubble parameter measurements, indicated that 
$H_0 t_0 = (H_0/70~{\rm km/s/Mpc})(t_0/14~{\rm Gyr}) \simeq 1$ or 
larger, which requires an epoch during which $a(t)$ grows 
as fast or faster than $t$, i.e., accelerated expansion (see Fig. \ref{fig:scale}a). 
As a result of this combination of factors, by the mid-1990's, a model with a dominant 
form of dark energy was recognized as a good match to much of 
the cosmological data \citep{Frieman_PNGB,Krauss_Turner,JPO_PJS}. 

On the other hand, the cosmological constant had a troubled 
history in the 20th century. Beginning with Einstein, 
it had been introduced to explain apparent observations 
that subsequently either evaporated or were explained on other 
grounds. These episodes included the preponderance of quasars 
around $z \sim 2$ in the late 1960's and the Hubble diagram 
of brightest-cluster elliptical galaxies in the mid-70's.  
Based on this history, healthy early skepticism of the supernova 
results was warranted. However, 
we now have independent, robust lines of evidence for 
cosmic acceleration from multiple sources: type Ia supernovae, 
the cosmic microwave background, and large-scale structure, among others. 
Moreover, as new data has accumulated, the evidence has strengthened.

\section{The Discovery of Cosmic Acceleration}

In this section we overview the evidence for the accelerating 
Universe that accrued in the late 1990's and early 2000's. 

\subsection{Type Ia Supernovae}

Supernovae were first studied systematically by Zwicky and 
collaborators in the 1930's. It was soon recognized that these 
luminous outbursts can 
be classified into different types based on their light curves 
and optical spectra. Type I supernovae were observed to 
exhibit similar light curves to each other and displayed no 
hydrogen in their spectra. Type II supernovae, identified by 
their strong hydrogen lines, showed much larger variability in 
their light curves. Type I supernovae were later subclassified into 
type Ia, which show silicon lines, type Ib, which show helium 
and no silicon, and type Ic, which show neither. 

Although this classification is purely empirical, the type Ia 
supernovae appear to be a physically distinct class from the 
other supernova types. SNe Ia are the thermonuclear explosions 
of white dwarf stars that are accreting mass from a binary 
companion and approaching the Chandrasekhar mass. The other supernova 
types arise from the core collapse of evolved massive stars. 

Type Ia supernovae appear quite homogeneous in their observational 
properties: they show similar spectral features, rise 
times of $15$ to $20$ days, and decay times of months. 
SN Ia light curves are powered by the radioactive decays of ${}^{56}$Ni (at
early times) and ${}^{56}$Co (after a few weeks), produced in the
thermonuclear explosion 
\citep{Hillebrandt_00}.  The peak luminosity is determined primarily by the
mass of ${}^{56}$Ni produced in the explosion \citep{Arnett_82}: if the white
dwarf is fully burned, one expects $\sim 0.6 M_\odot$ of ${}^{56}$Ni to be
produced.  As a result, although the detailed mechanism of SN Ia explosions 
remains uncertain \citep[e.g.,][]{hoflich04,plewa},
SNe Ia are expected and observed to have
similar peak luminosities.  

In fact, SNe Ia are not precisely 
standard candles, with a $1\sigma$ spread of
order 0.3 mag in peak $B$-band luminosity.  
However, work in the early 1990's \citep{Phillips_93}
established an empirical correlation between SN Ia peak brightness and the
rate at which the luminosity declines with time after peak: intrinsically
brighter SNe Ia decline more slowly. After correcting for this correlation,
SNe Ia turn out to be excellent ``standardizable'' candles, with a dispersion
of about 15\% in peak brightness.

To put the supernova observations in context, we pause to review 
how distances are defined and measured in cosmology.

\subsubsection{Cosmological Distances}

The spacetime metric for the homogeneous, isotropic FRW cosmology can 
be written in either of two useful forms:

\begin{eqnarray} 
ds^2 &=& c^2dt^2 - a^2(t) \left[ dr^2/(1-kr^2) +r^2 \left(d\theta^2 +\sin^2 \theta d\phi^2\right) \right] \nonumber \\
&=& c^2 dt^2 - a^2(t)\left[d\chi^2 + S_k^2(\chi)\left(d\theta^2 +\sin^2 \theta d\phi^2\right) \right]   ~,
\label{eq:metric}
\end{eqnarray}

\noindent where $k$ is the spatial curvature index, $\theta$, $\phi$ are the usual angular coordinates in a spherical 
coordinate system, and $r=S_k(\chi)=\sinh(\chi), \chi, \sin(\chi)$ for 
$k=-1, 0, +1$. Along a radial null geodesic (light ray), 
$ds^2=d\theta^2=d\phi^2=0$, so $c dt=a d\chi$. As a result, 
the comoving distance is given by 

\begin{equation}
\chi = \int {c dt \over a} = \int {c dt \over a da} da = c \int {da \over a^2 H(a)}~.
\label{eq:chi}
\end{equation}

\noindent Without loss of generality we can 
set $a(t_0)\equiv a_0=1$, so that $a=1/(1+z)$, and we have 
$da = -(1+z)^{-2} dz = -a^2 dz$. We therefore derive 
a simple relation between redshift and comoving distance,  
$c dz = -H(z) d\chi$. 

To compute the {\it luminosity distance}, consider a source $S$ at the origin emitting 
light at time $t_1$ into solid angle $d\Omega$ that is received by observer $O$ at coordinate 
distance $r$ at time $t_0$ who has a detector of area $A$. The proper area of the detector 
is given by the FRW metric, $A=a_0 r d\theta a_0 r \sin \theta d\phi =a_0^2 r^2 d\Omega$. 
A unit-area detector at $O$ thus subtends a solid angle $d\Omega = 1/a_0^2 r^2$ at $S$. 
The power emitted into $d\Omega$ by a source of luminosity $L$ is $dP=L d\Omega/4 \pi$, and the energy flux received by 
$O$ per unit area is thus $f=L d\Omega/4\pi = L/4\pi a_0^2 r^2$. However, the expansion of the Universe 
reduces the received flux due to two effects: (i) the photon energy redshifts, 
$E_\gamma(t_0)=E_\gamma(t_1)/(1+z)$; (ii) photons emitted at time intervals $\delta t_1$ 
arrive at larger time intervals given by $\delta t_0/\delta t_1 =a(t_0)/a(t_1)=1+z$: this 
time dilation factor can be derived from the constancy of the comoving distance of Eqn. \ref{eq:chi}, 
i.e., by setting $\int_{t_1}^{t_0}dt/a(t)=\int_{t_1+\delta t_1}^{t_0+\delta t_0}dt/a(t)$ 
and rewriting the integration limits. Including these two factors, we can write the 
flux as

\begin{equation}
f={L d\Omega \over 4\pi}={L \over 4\pi a_0^2 r^2(1+z)^2} \equiv {L\over 4\pi d_L^2}~,
\label{eq:flux}
\end{equation}

\noindent where the last equality defines the luminosity distance $d_L$ in terms of the usual 
inverse-square law, 

\begin{equation}
d_L = r(1+z) = c(1+z) {|\Omega_k|}^{-1/2} S_k \left(\int {|\Omega_k|}^{1/2}{da \over H_0 a^2 (H(a)/H_0)}\right)~,
\label{eq:dL}
\end{equation}

\noindent where $\Omega_k = 1-\Omega_0 = 1 - \Omega_m- \Omega_{DE}$. 
For a general dark energy model with equation of state parameter $w(z)$, the Hubble expansion 
rate can be written as

\begin{equation}
{H^2(z)\over H_0^2}=\Omega_m(1+z)^3 + \Omega_{DE} \exp \left[3\int(1+w(z))d\ln(1+z)\right] + 
\Omega_k(1+z)^2~. 
\label{eq:Hz}
\end{equation}

\noindent For the case of the cosmological constant, $w=-1$, 
this can be rewritten as 

\begin{equation}
{H^2(a)\over H^2_0}=\Omega_m a^{-3}+\Omega_\Lambda+\Omega_ka^{-2}~,
\label{eq:HL}
\end{equation}

\noindent and the luminosity distance becomes

\begin{equation}
d_L(z;\Omega_m, \Omega_\Lambda) = {c(1+z)\over H_0}{|\Omega_k|}^{-1/2} S_k \left( \int_{1/(1+z)}^1 
{|\Omega_k|}^{1/2} {da \over a^2[\Omega_m a^{-3}+\Omega_\Lambda 
+\Omega_k a^{-2}]^{1/2}}\right)~.
\label{eq:dLL}
\end{equation}

\noindent Another important special case is a flat Universe ($\Omega_k=k=0$) and dark energy with 
$w=$ constant, independent of redshift. In this case,

\begin{equation}
{H^2(z)\over H^2_0}=(1-\Omega_{DE})(1+z)^3+\Omega_{DE}(1+z)^{3(1+w)}~,
\label{eq:Hz2}
\end{equation}

\noindent and the luminosity distance is given by 

\begin{equation}
d_L(z;\Omega_{DE},w)=\chi(1+z) = {1+z \over H_0}\int{1+\Omega_{DE}[(1+z)^{3w}-1]^{-1/2}\over (1+z)^{3/2}}dz~.
\label{eq:dLL2}
\end{equation}

\begin{figure}
\centerline{\psfig{file=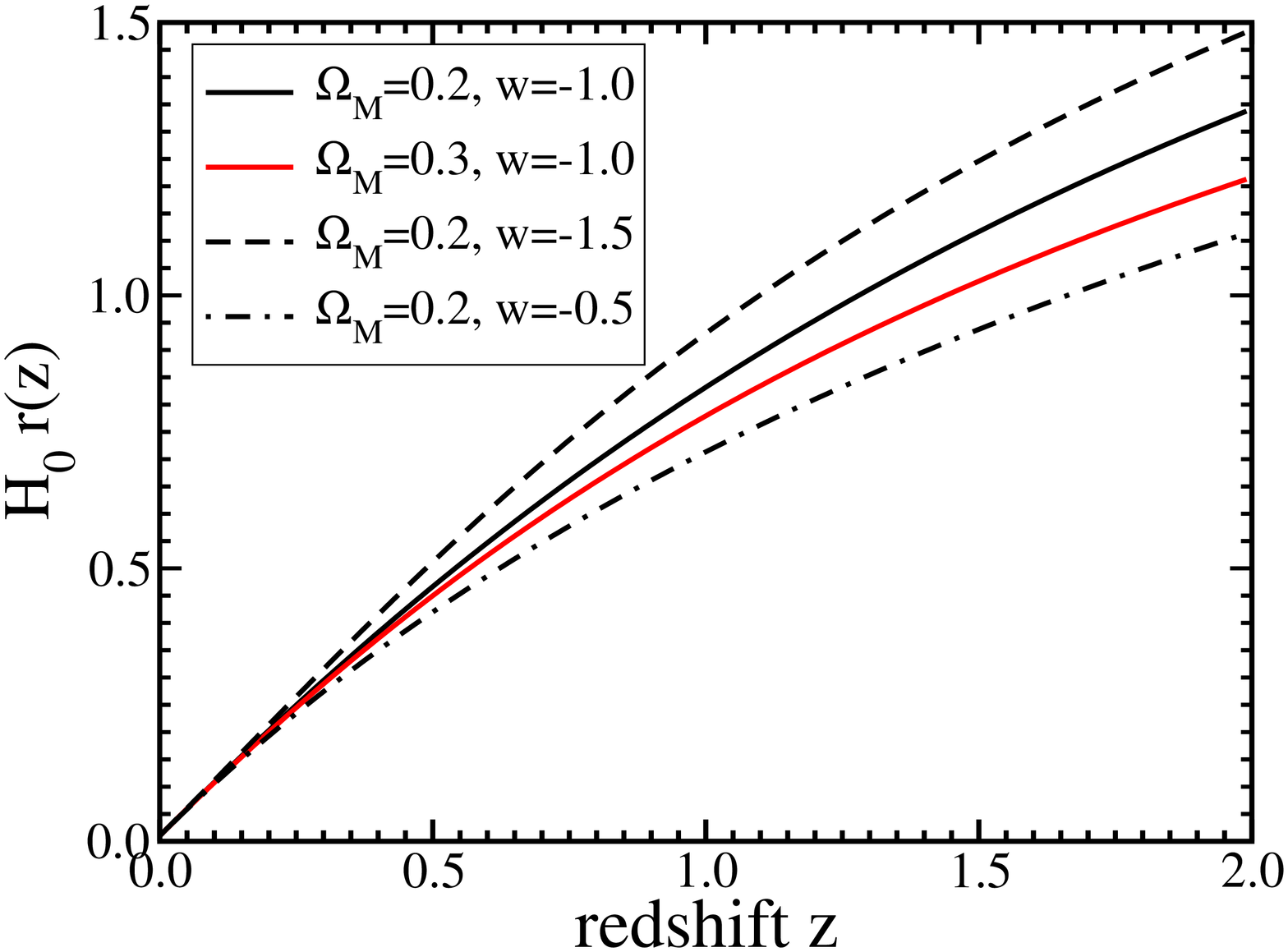,height=2.4in,angle=-90}
\psfig{file=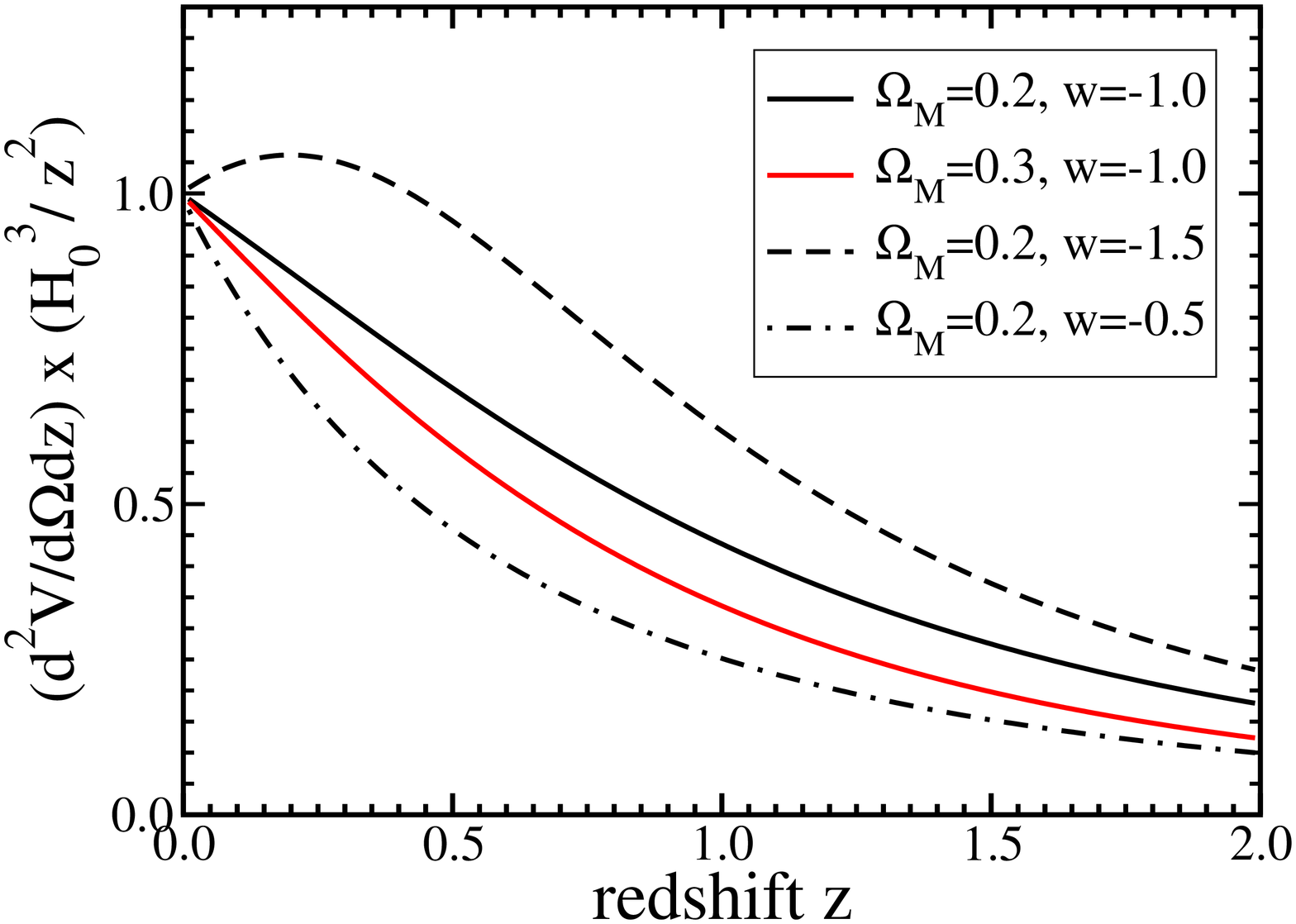,height=2.4in,angle=-90}}
  \caption{Left: Distance vs. redshift in a flat Universe with different values of the 
cosmological parameters $\Omega_m$ and $w$. Right: volume element vs. redshift 
for same models. From \citet{fth08}.}
\label{fig:rz}
\end{figure}

\noindent Note that the product $H_0 d_L$ is independent of the Hubble parameter $H_0$.

The absolute and apparent magnitudes are logarithmic measures of luminosity and flux:
$M_j = -2.5 \log(L_j)+c_1$, $m_i=-2.5\log(f_i)+c_2$, where $i(j)$ denotes the observed(rest-frame) 
passband. Luminosity distance measurements are then conveniently given in 
terms of the {\it distance modulus}, 

\begin{equation}
\mu \equiv m_i - M_j = 2.5 \log(L_j/f_i)= 5 \log[H_0 d_L(z;\Omega_m,\Omega_{DE}, w(z))]-5\log H_0 
+ K_{ij}(z) ~,
\label{eq:mu}
\end{equation}

\noindent where $K_{ij}$ is the redshift-dependent K-correction that accounts for the 
redshifting between the observed and emitted passbands and depends upon the spectral 
energy distribution of the source. For a population of standard candles (fixed $M_j$) with 
known spectra ($K_{ij}$), measurements of $\mu$ vs. $z$, the Hubble diagram, constrain 
cosmological parameters. The parameter dependence of the distance vs. redshift is 
shown in the left panel of Fig. \ref{fig:rz}. The right panel shows the comoving 
volume element, $d^2 V/dz d\Omega = r^2(z)/H(z)$.

If $M_j$ is known, then from measurement of $m_i$ and knowledge of the 
spectrum we can infer the {\it absolute} distance to an object at redshift $z$; we can  
thereby determine $H_0$, since $d_L \simeq cz/H_0$ for $z \ll 1$. If $M_j$ is unknown, then 
from measurement of $m_i$ we can infer the distance to an object at redshift $z_1$ {\it relative} 
to an object at redshift $z_2$, $m_1 - m_2 = 5\log(d_1/d_2)+K_1-K_2$. For supernovae, we 
typically measure relative distances, using low-redshift supernovae to vertically anchor 
the Hubble diagram, i.e., to approximately determine the quantity $M-5\log H_0$.

\subsubsection{SN Discovery}

\begin{figure}
\centerline{\psfig{file=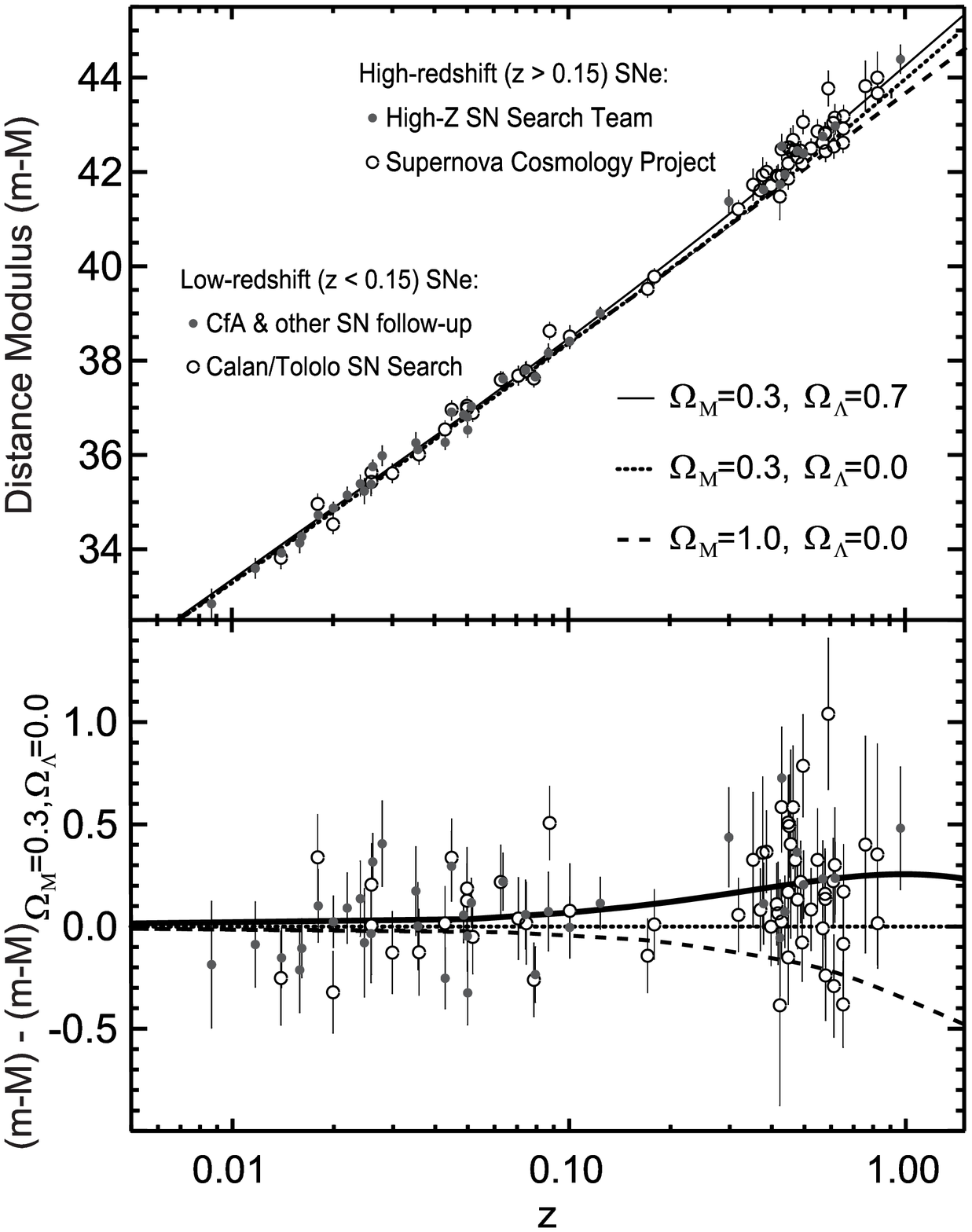,height=3.5in}
\psfig{file=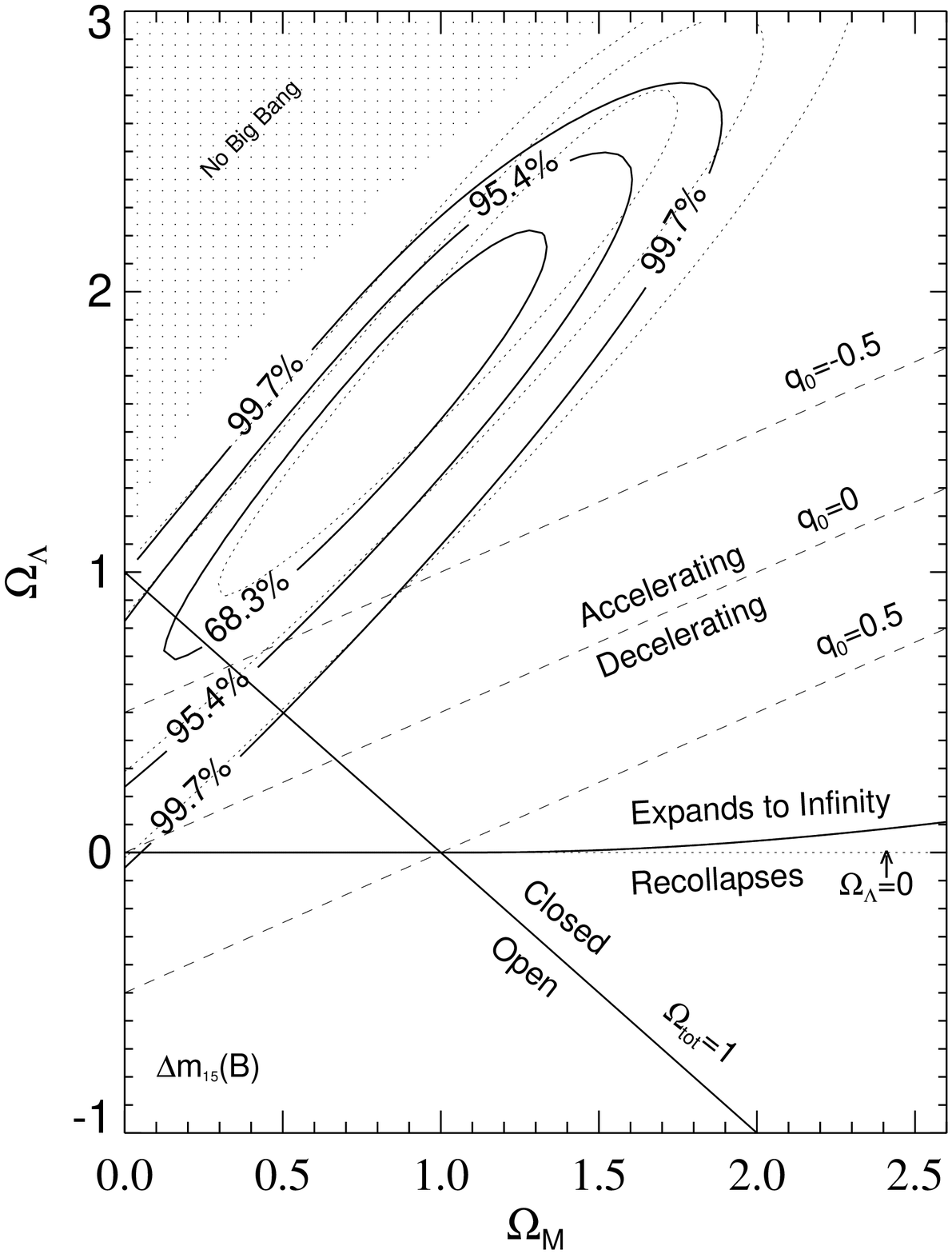,height=3.7in}}
  \caption{Left panel (a): Discovery data: Hubble diagram of SNe Ia measured by the SCP and HZT. Bottom 
panel shows residuals in distance modulus relative to an open universe with 
$\Omega_0=\Omega_m=0.3$. Figure adapted from \cite{Riess_2000,Perlmutter_Schmidt}, based on \cite{riess98,perlmutter99}.
Right panel (b): constraints on $\Omega_m$ and $\Omega_\Lambda$ from the HZT data \cite{riess98}.
}
\label{fig:PS}
\end{figure}

The recognition in the 1990's that supernovae are standardizable candles, together with the availability 
of large mosaic CCD cameras on 4-meter class telescopes, stimulated two teams, the 
Supernova Cosmology Project (SCP) and the High-z SN Search Team (HZT), to measure the SN Ia Hubble 
diagram to much larger distances than was previously possible. Based on 
samples of tens of objects, both teams found that 
distant SNe are $\sim 0.25$ mag dimmer than they would be in a decelerating Universe, 
indicating that the expansion has been speeding up for the past 5 Gyr \citep{riess98,perlmutter99}; 
a compilation of the discovery data from the two teams is shown in Fig. \ref{fig:PS}a. Initially, 
these results were interpreted in terms of the cosmological constant model, Fig. \ref{fig:PS}b,  
using Eqn. \ref{eq:dLL}. The constraint region delineates the values of the parameters 
$\Omega_m, \Omega_\Lambda$ which combine to give similar luminosity distance estimates to 
$z \sim 0.5$. The results are also often interpreted in terms of the flat, constant $w$ 
model of Eqn. \ref{eq:dLL2}, as shown in Fig. \ref{fig:kowalski} below.

\subsection{Cosmic Microwave Background Anisotropy}

\begin{figure}
 \includegraphics[height=.35\textheight]{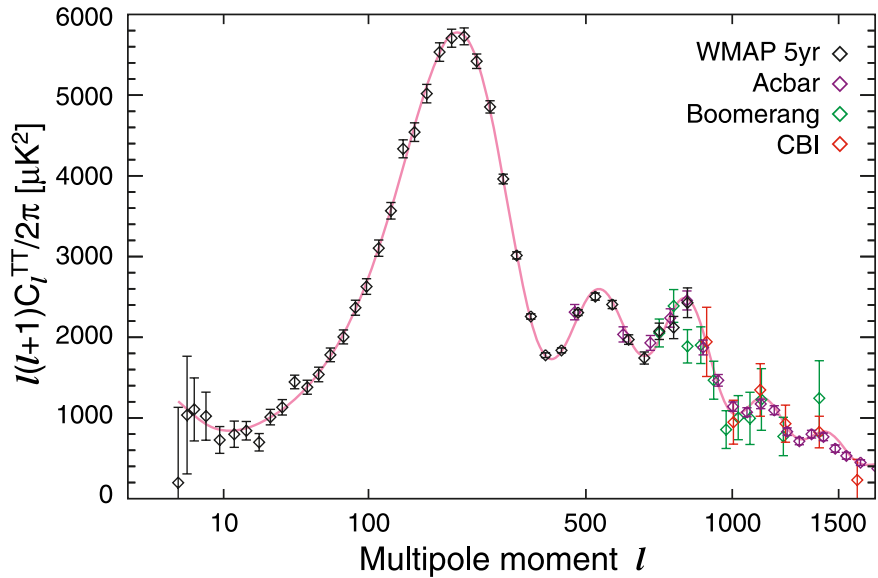}
  \caption{CMB temperature anisotropy angular power spectrum measurements from WMAP, 
Acbar, Boomerang, and CBI. From \citet{dunkley08}.
}
\label{fig:CMB}
\end{figure}

The second important piece of evidence in favor of cosmic acceleration came from 
the CMB anisotropy. As noted above, the CMB carries the imprint of processes in 
the photon-baryon fluid around the time of recombination, when the photons 
last scattered with baryons, at $z_{ls} =1089$. 
CMB maps, such as those made most recently 
by the Wilkinson Microwave Anisotropy Probe (WMAP), 
show the strongest temperature fluctuations on a characteristic angular scale of about 1 degree. 
This is the angular scale subtended by the sound horizon, $s=c_s t_{ls}$, the 
distance that gravity-driven sound waves in the photon-baryon fluid can travel before 
last scattering. More precisely, 

\begin{equation}
s= \int_{0}^{t_{ls}} c_s (1+z) dt = \int_{z_{ls}}^{\infty} {c_s \over H(z)}dz~,
\label{eq:sound}
\end{equation}

\noindent where the sound speed $c_s$ is determined by the ratio of the baryon and photon
energy densities, $c_s = [3(1+3\rho_b/4\rho_\gamma)]^{-1/2}$. 
Before recombination, the sound speed is relativistic, 
about $c_s \simeq 0.57c$, 
due to the large pressure provided by the CMB photons. 

While $s$ is fixed 
primarily by $\Omega_\gamma$ and $\Omega_b$, the 
angular scale subtended by $s$ is determined primarily by the spatial curvature 
of the Universe: for $\Omega_0 >1$ ($\Omega_0<1$), the angular scale is larger (smaller) 
than it is in a flat Universe. In 2000-2001, the Boomerang, DASI, and MAXIMA experiments 
\citep{Boom,DASI,MAXIMA} 
reported robust detections of the first acoustic peak in the CMB temperature 
anisotropy angular power spectrum at an angular scale of about 1 degree, as 
expected for a nearly flat Universe. In a plot of $\Omega_\Lambda$ vs. $\Omega_m$ 
such as Fig. \ref{fig:PS}b, the CMB constraints thus lie near the $\Omega_0=1$ 
line, nearly orthogonal to the supernova constraints; more precisely, the 
CMB degeneracy line is approximately $1-\Omega_0=-0.3+0.4\Omega_\Lambda$, 
as shown below in Fig. \ref{fig:kowalski}a. Together, the 
SN and CMB constraints point to a best-fit model with $\Omega_\Lambda \simeq 0.75$ and 
$\Omega_m \simeq 0.25$. The CMB constraints have been strengthened considerably 
by recent results from WMAP and from ground-based experiments that probe 
smaller angular scales. A recent compilation of CMB anisotropy results is 
shown in Fig. \ref{fig:CMB}, clearly showing the pattern of acoustic oscillations.  
The WMAP 5-year data constrains the distance to last scattering as \citep{komatsu08}

\begin{equation}
R=(\Omega_m H_0^2)^{1/2} \int_{0}^{z_{ls}}{dz \over H(z)}=1.715 \pm 0.021~.
\label{eq:Rcmb}
\end{equation}

\noindent The resulting constraints on cosmological parameters are 
shown in Fig. \ref{fig:kowalski}. The CMB constraints on the dark energy equation of state parameter $w$ 
are comparatively 
weak, due to a large degeneracy between $w$ and $\Omega_{DE}$ in determining the 
angular diamater distance to the last-scattering surface. Nevertheless, 
the positions and amplitudes of the acoustic peaks in Fig. \ref{fig:CMB} encode 
a wealth of cosmological information. CMB measurements 
are extremely important for present and future dark energy probes since they strongly 
constrain a variety of cosmological parameters. The upcoming Planck mission is expected 
to constrain a number of (non-dark energy) cosmological parameters at the $\sim 1$\% level.

\subsection{Large-scale Structure: Baryon Acoustic Oscillations}

\begin{figure}
 \includegraphics[height=.35\textheight]{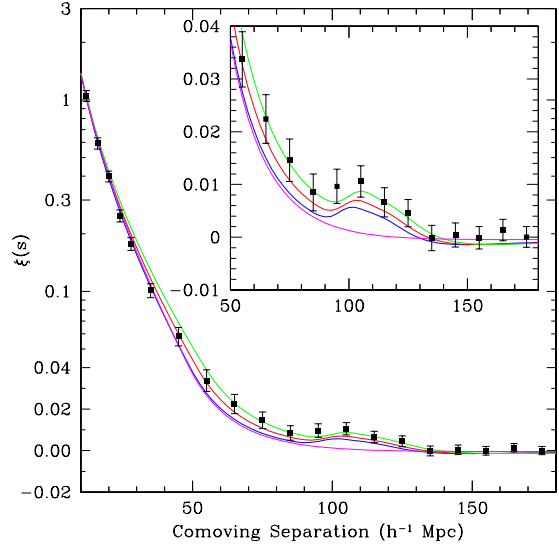}
  \caption{Correlation function for SDSS luminous red galaxies, showing 
the expected bump due to BAO on large scales. From \citet{SDSS_BAO}.
}
\label{fig:sdsslrg}
\end{figure}

The baryon acoustic oscillations (BAO), so prominent in the CMB anisotropy, leave a more subtle imprint 
in the large-scale distribution of galaxies, a bump in the two-point correlation function 
$\xi(r)$ at a scale $r \sim 110 h^{-1}$ Mpc. This is roughly the distance at which an outgoing spherical 
sound wave in the photon-baryon fluid stalls when the 
photons and their associated pressure 
decouple from the baryons at the time of last scattering. The sound waves 
remain imprinted in the baryon distribution and, through gravitational 
interactions, in the dark matter distribution as well. However, 
since $\Omega_b/\Omega_m \simeq 
1/6$, dark matter dominates the growth of structure, and the imprint of 
baryon oscillations in the galaxy distribution is relatively small.

Measurement of the BAO signature in the correlation 
function of luminous red galaxies in the Sloan Digital Sky Survey (SDSS), shown 
in Fig. \ref{fig:sdsslrg}, has constrained 
the distance to the median redshift $z_1=0.35$ of this sample to a precision of 5\% \citep{SDSS_BAO}.
The BAO measurements constrain several different parameters \citep{SDSS_BAO,percival07}, depending 
on whether and how information from the CMB is used. \citet{SDSS_BAO} show that they 
constrain the combination of angular diameter distance, Hubble parameter, and $\Omega_m$ 
given by

\begin{eqnarray}
A(z_1;w,\Omega_m,\Omega_{DE}) &=& \sqrt{\Omega_m}\left({H_0\over H(z_1)}\right)^{1/3} \left[{|\Omega_k|^{-1/2} 
\over z_1} S_k \left(|\Omega_k|^{1/2} \int_0^{z_1} dz {H_0 \over H(z)}\right)\right]^{2/3} \nonumber \\
&=& 0.469 \pm 0.017~.
\label{eq:Abao}
\end{eqnarray}

\noindent Since this quantity scales with redshift and cosmological parameters in a manner 
different from the luminosity distance, its measurement carves out a different likelihood 
region in the space of cosmological parameters.
The resulting constraints in the $w, \Omega_m$ plane are shown in Fig. \ref{fig:kowalski}. This figure 
demonstrates the robustness of the current results: although SN, CMB, and LSS 
are complementary, one can drop any one of them and still find strong 
evidence in favor of cosmic acceleration. For evidence from the CMB and LSS 
alone, see, e.g., \citep{SDSS_LRG}.

\section{Continuing Evidence for Cosmic Acceleration}

While CMB and LSS measurements 
independently strengthened the evidence for an accelerating Universe, 
subsequent supernova observations have reinforced the original results, and 
new evidence has accrued from other observational probes. Here we briefly 
review these recent developments and discuss the current status of our knowledge 
of dark energy. 

\subsection{Recent Supernova Observations}

\begin{figure}
\includegraphics[trim = 0mm 0mm 43mm 0mm, clip,height=3.6in]{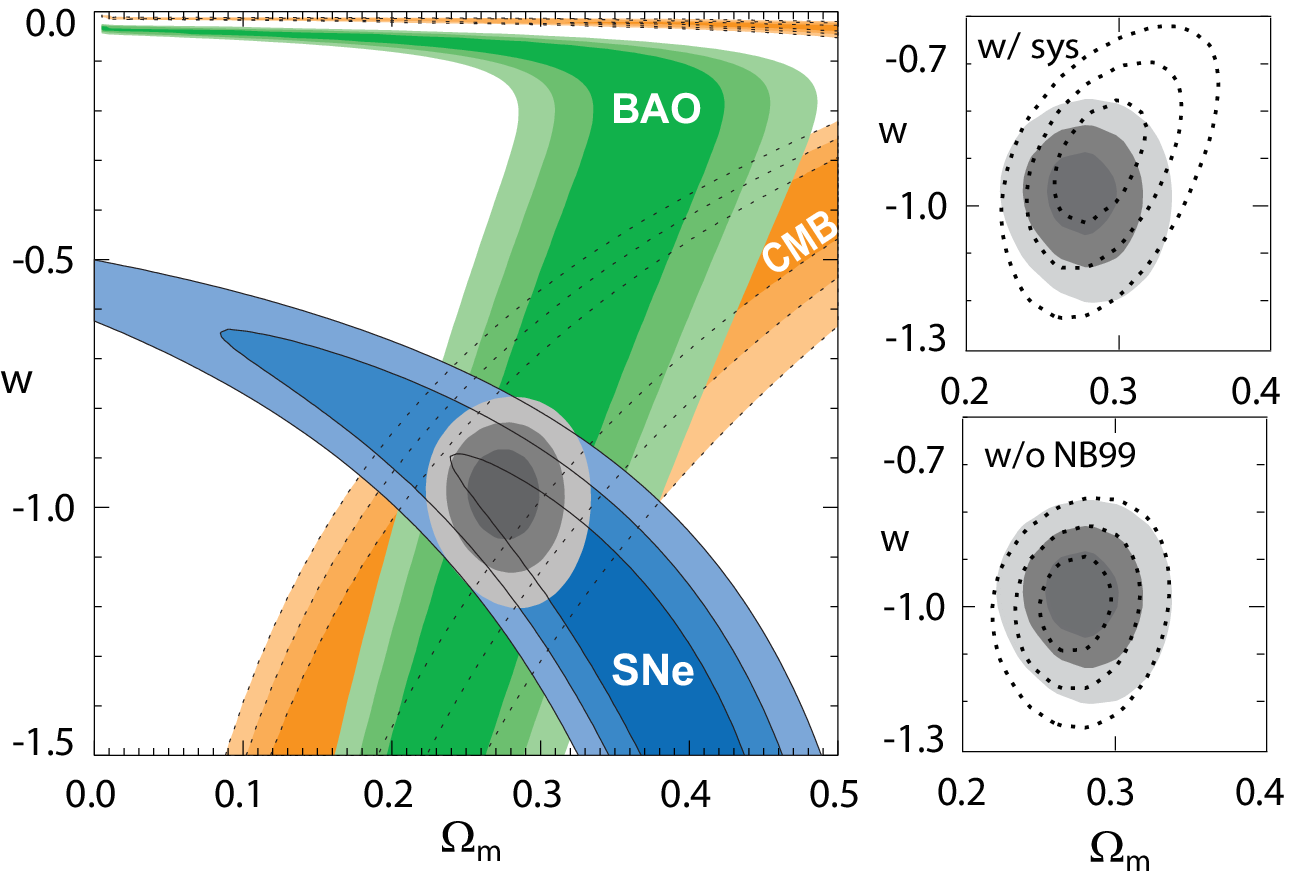}
\includegraphics[height=3.6in]{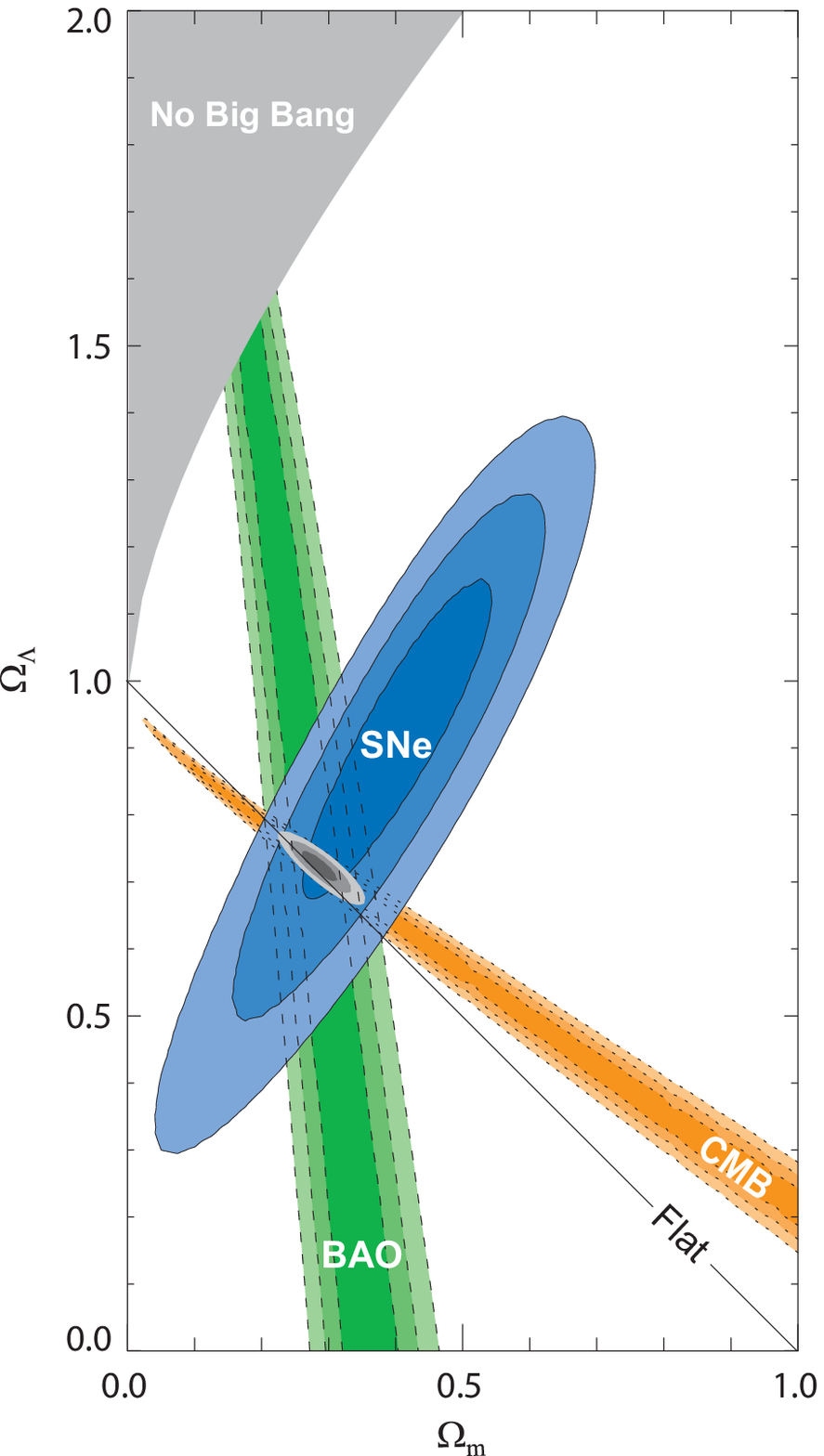}
  \caption{Left panel (a): recent constraints from SNe, CMB anisotropy (WMAP5), and 
large-scale galaxy correlations (SDSS BAO) upon the cosmological parameters 
$w, \Omega_m$ (for a flat Universe with constant $w$); Right panel (b): constraints upon 
$\Omega_m$ and $\Omega_\Lambda$ for the cosmological constant model ($w=-1$). 
Statistical 
errors only are shown. 
From \citet{kowalski08}.
}
\label{fig:kowalski}
\end{figure}

A number of
concerns were raised about the robustness of the first SN evidence 
for acceleration, e.g., it was suggested that distant SNe could 
appear fainter due to extinction by hypothetical 
grey dust rather than acceleration 
\citep{Drell00,Aguirre99}.  Over the intervening decade, the supernova
evidence for acceleration has been strengthened by results from a series of SN
surveys. Observations with the Hubble Space Telescope (HST) have provided
high-quality light curves \citep{Knop03} and have extended SN measurements to
redshift $z\simeq 1.8$, providing evidence for the expected earlier epoch of
deceleration and disfavoring dust extinction as an alternative 
explanation to acceleration \citep{Riess_01,Riess_04,Riess_06}. 

Two large ground-based surveys, the
Supernova Legacy Survey (SNLS) \citep{SNLS} 
and the ESSENCE survey 
\citep{Miknaitis_07,WoodVasey_07}, have been using 4-meter telescopes to measure 
light curves for several hundred SNe Ia over the redshift range $z \sim
0.3-0.9$, with large programs of spectroscopic follow-up on 6- to 10-m
telescopes. They have each reported results from a fraction of the total 
data collected, with more to follow. The quality and quantity 
of the distant SN data are now vastly superior to what was available in 
1998, and the evidence for acceleration is correspondingly more 
secure. An example of a recent analysis using this larger, more recent 
combination of supernova data sets is shown in Fig. \ref{fig:kowalski} \cite{kowalski08}.
In combination with CMB and LSS results (see below), the SN data 
constrain the equation of state parameter $w$ to an uncertainty $\sim 0.08-0.15$ 
for a flat Universe and constant $w$, 
with a central value consistent with the cosmological constant, $w=-1$. 
The range of uncertainty above reflects different assumptions in the literature 
about the size of systematic errors. On 
the other hand, if we drop the assumption of non-evolving $w$, which is not 
well motivated if $w \neq -1$, then the current 
constraints are considerably weaker.

\subsection{Systematic Errors in SN Distance Estimates}

With the substantial advances in the number of 
supernova distance measurements in recent years and the consequent decline 
in statistical errors, systematic errors have come into sharper focus as a 
limiting factor. 
The major systematic concerns for supernova distance measurements are errors in 
correcting for host-galaxy extinction and uncertainties in the intrinsic colors 
of supernovae; luminosity evolution; and selection bias in the low-redshift sample. 
Even with multi-band observations, the combination of photometric errors, variations 
in intrinsic SN Ia colors, and uncertainties and likely variations in host-galaxy 
dust properties lead to significant uncertainties in distance estimates. Observations 
that extend into the rest-frame near-infrared, such those being carried out by the 
Carnegie Supernova Project, offer promise in controlling dust-extinction systematics 
since their effects are much reduced at long wavelengths.

With respect to luminosity evolution, there is evidence 
that SN peak luminosity correlates with
host-galaxy type \citep[e.g.,][]{JRK_07}, and that the mean host-galaxy environment,
e.g., the star formation rate, evolves strongly with look-back time.  However,
brightness-decline-corrected SN Ia Hubble diagrams are consistent between
different galaxy types, and since the nearby Universe spans the range of
galactic environments sampled by the high-redshift SNe, one can measure
distances to high-redshift events by comparing with low-redshift analogs.
While SNe provide a number of correlated observables (multi-band light curves
and multi-epoch spectra) to constrain the physical state of the system,
insights from SN Ia theory will likely be needed to determine if they are collectively
sufficient to constrain the mean peak luminosity at the percent level required 
for future dark energy missions 
\citep{hoflich04}.  

\begin{figure}
\centerline{\psfig{file=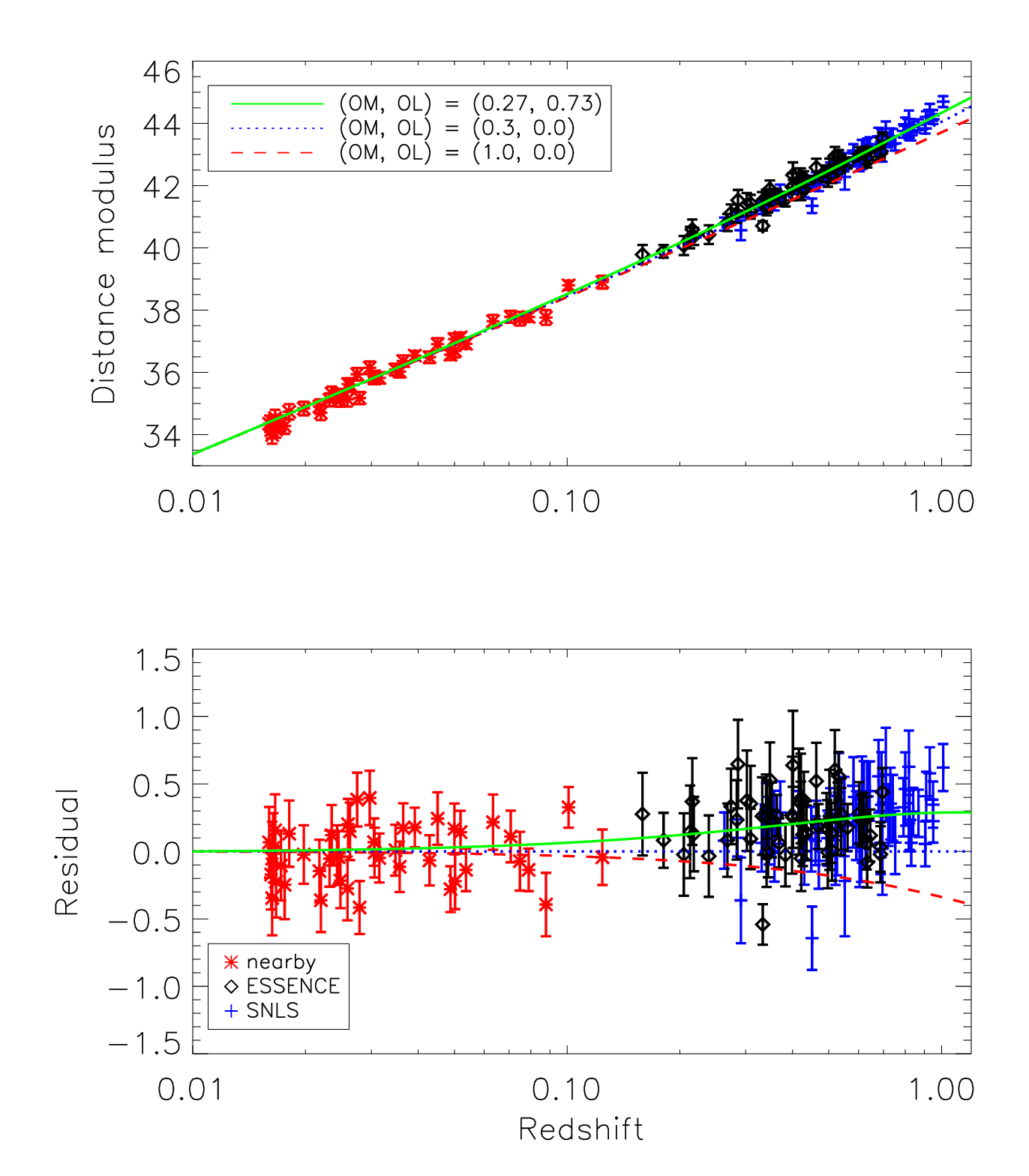,height=3.4in}
\psfig{file=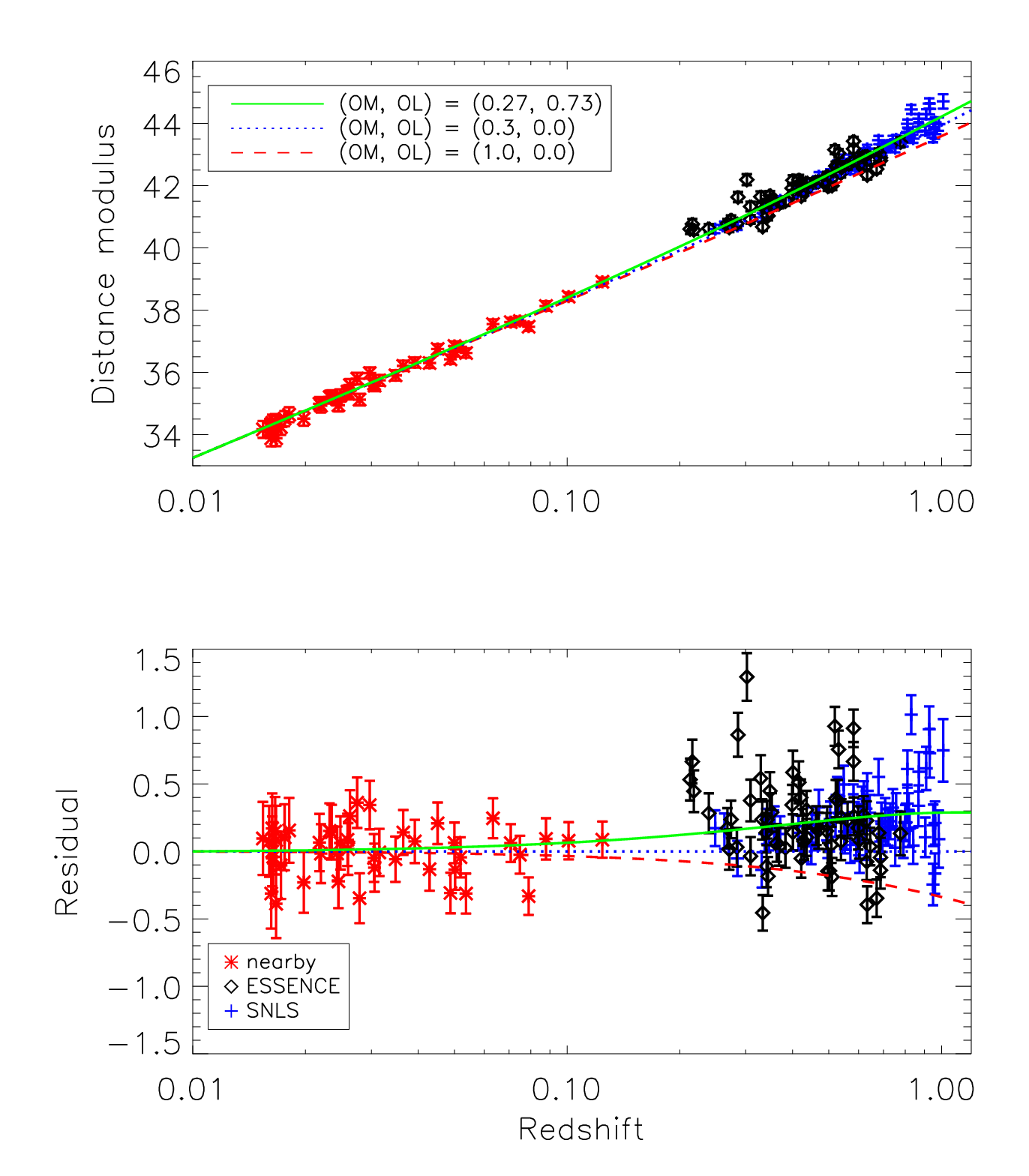,height=3.4in}}
  \caption{Hubble diagram for low-redshift (red), SNLS (blue), and ESSENCE (black) 
supernovae using the MLCS2k2 distance estimator (left panel) and the SALT 
fitter (right panel). Lower panels show residuals from an open model with 
$\Omega_m=0.3, \Omega_\Lambda=0$. From \citet{WoodVasey_07}.
}
\label{fig:saltmlcs}
\end{figure}

Finally, there is concern that the low-redshift SNe currently used to anchor 
the Hubble diagram and that serve as templates for fitting
distant SN light curves are a relatively small, heterogeneously selected
sample and that correlated large-scale peculiar velocities induce 
larger distance errors than previously estimated \citep{hui_06}.  
This situation should improve in the near future once results are
collected from low-redshift SN surveys such as the Lick Observatory Supernova
Search (LOSS), the Center for Astrophysics Supernova project, the Carnegie Supernova
Project, the Nearby Supernova Factory, and the SDSS-II
Supernova Survey. Over the course of three 3-month seasons, the 
latter survey discovered and measured multi-band light curves 
for nearly 500 spectroscopically confirmed SNe Ia in the redshift range $z \sim 0.05-0.4$, 
filling in the ``redshift desert'' between low- and high-redshift samples that is
evident in Fig. \ref{fig:saltmlcs} \citep{frieman08,sako08}.

One illustration of systematic error concerns is provided by Fig. \ref{fig:saltmlcs}, 
which shows the Hubble diagram for the same published low-redshift, ESSENCE, and SNLS 
data analyzed using two different light-curve fitters, the multi-color light-curve 
shape (MLCS) method \citep{JRK_07} and SALT \citep{Guy1,Guy2}. One sees significant 
differences in distance estimates, and the best-fit values of $w$, using the 
SNe plus BAO constraints, differ by about 0.11. These fitters take 
different approaches to SN distance estimation. MLCS is based on $UVBRI$ 
rest-frame light-curve templates derived from well-observed nearby supernovae. 
It incorporates a brightness vs. light-curve shape correlation and assumes 
that color variations not associated with the brightness-shape relation are 
due to dust extinction in the host galaxy. Host-galaxy dust is modeled using 
a derived extinction vs. wavelength relation for the Milky Way, and a prior 
on the amplitude of the dust extinction is imposed, again based on observations 
of the low-redshift supernovae. SALT, by contrast, begins with rest-frame spectral templates
that are synthesized to produce broad-band model photometry. It incorporates 
a brightness-shape correlation and an empirical color variation that is not required 
to emulate the effects of dust but that appears to match the wavelength dependence of 
dust extinction reasonably well. In particular, there are no explicit assumptions 
about dust and no associated priors. However, the translation of light-curve shape 
and color variations into luminosity and therefore distance variations is controlled 
by two global parameters that are determined in a simultaneous fit of model and 
cosmological parameters to the Hubble diagram. As a result, SALT does not yield 
cosmology-independent distance estimates for each supernova, and SALT SN distances 
derived in the context of a particular cosmology parametrization (e.g., 
$k=0$ and constant $w$) should not be applied to constrain other models.

Accounting for systematic errors, precision measurement of $w$ and particularly 
of its evolution with redshift will require a few thousand 
SN Ia light curves out to redshifts $z\sim 1.5$ to be measured with
unprecedented precision and control of systematics \citep{frieman_03}.  
For redshifts $z>0.8$, this will 
require going to space to minimize photometric errors, to obtain uniform
light-curve coverage, and to observe in the near-infrared bands to capture the 
redshifted photons.

\subsection{Corroborating Evidence for Acceleration}

A number of observations made in the last several years have provided additional 
evidence for cosmic acceleration. We highlight some of the major developments here.

{\bf Integrated Sachs-Wolfe effect:} The presence of dark energy affects the large-angle anisotropy of the CMB (the
low-$\ell$ multipoles). This Integrated Sachs-Wolfe (ISW) effect 
arises due to the differential redshifts of photons as they pass through
time-changing gravitational potential wells, and it leads to a small correlation
between the low-redshift matter distribution and the CMB temperature anisotropy.  This
effect has been observed in the cross-correlation of the CMB with
galaxy and radio source catalogs
\citep{Boughn_Crittenden,Fosalba_Gaztanaga,Afshordi_Strauss,Scranton_ISW}. This 
signal indicates that the Universe is not described by the Einstein-de Sitter
model ($\Omega_{m}=1$), a reassuring cross-check.

{\bf Weak gravitational lensing} \citep{Schneider_review,Munshi_review}, the small,
correlated distortions of galaxy shapes due to gravitational lensing by
intervening large-scale structure, is a powerful technique for mapping dark
matter and its clustering. Detection of this cosmic shear signal was first
announced by four groups in 2000 \citep{bacon00,kaiser00,vanW00,Witt00}.
Recent lensing surveys covering areas of order 100 square degrees have shed
light on dark energy by pinning down the combination $\sigma_8
(\Omega_{m}/0.25)^{0.6}\approx 0.85\pm 0.07$, where $\sigma_8$ is the rms
amplitude of mass fluctuations on the $8~h^{-1}$ Mpc scale
\citep{Jarvis_CTIO,Hoekstra,Massey}.  Since other measurements peg
$\sigma_8$ at $\simeq 0.8$, 
this implies that $\Omega_{m} \simeq 0.25$, consistent with
a flat Universe dominated by dark energy. In the future, weak lensing has the
potential to be a very powerful probe of dark energy
\citep{Huterer_thesis,Hu02}, as discussed below.

{\bf X-ray Clusters:} Measurements of the ratio of X-ray emitting gas to total mass in
galaxy clusters, $f_{\rm gas}$, also indicate the presence of dark energy.
Since galaxy clusters are the largest collapsed objects in the universe, the
gas fraction in them is presumed to be constant and nearly equal to the baryon
fraction in the Universe, $f_{\rm gas} \approx \Omega_{b}/\Omega_{m}$ 
(most of the
baryons in clusters reside in the gas).  The value of $f_{\rm gas}$ inferred
from observations depends on the observed X-ray flux and temperature as well
as the distance to the cluster.  Only the ``correct cosmology'' will produce
distances which make the apparent $f_{\rm gas}$ constant in redshift. Using
data from the Chandra X-ray Observatory, \citet{Allen_04,Allen_07} determined
$\Omega_\Lambda$ to a 68\% precision of about $\pm 0.2$, obtaining a value 
consistent with the SN data.

\begin{figure}
\centerline{\psfig{file=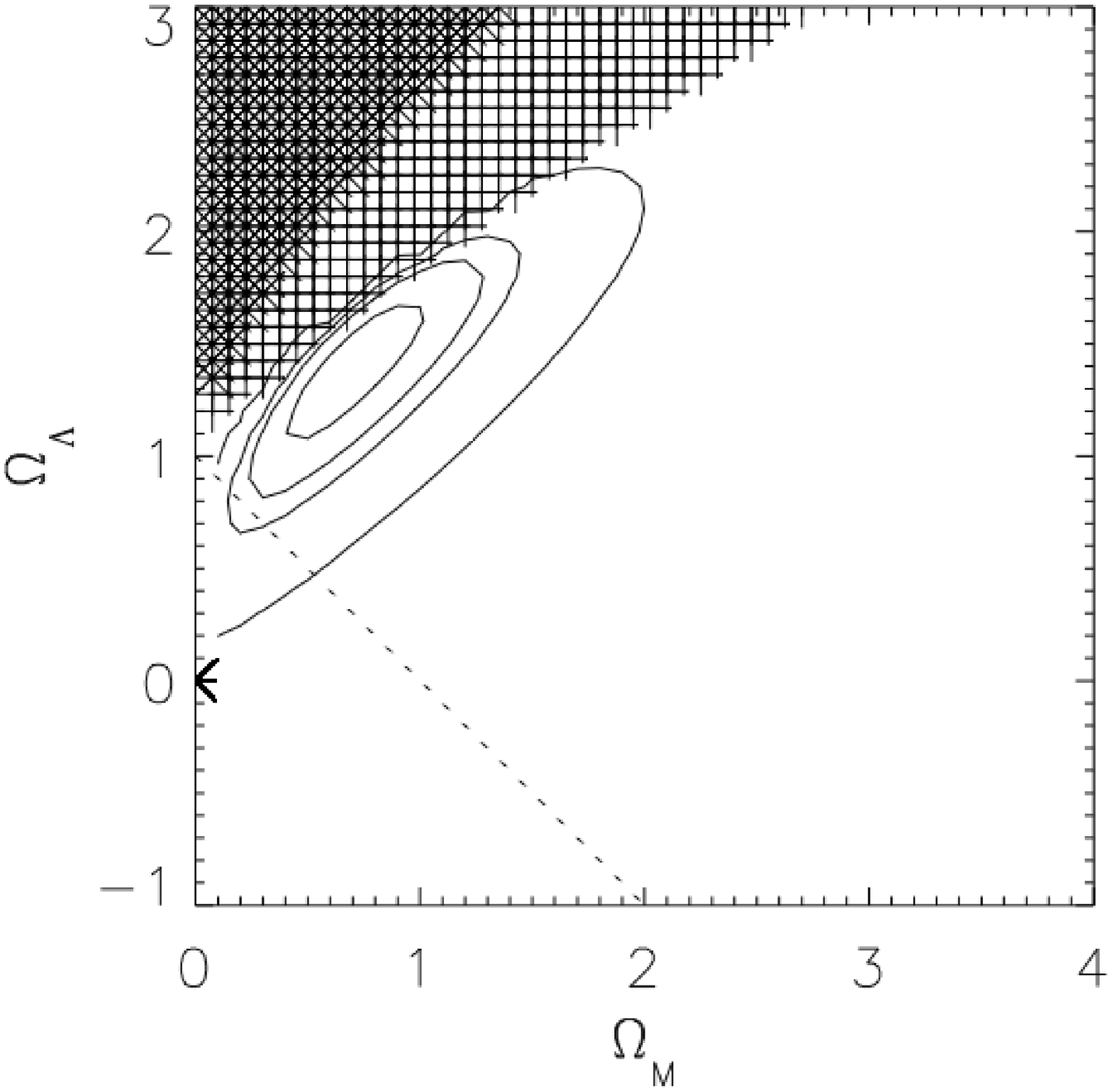,height=2.8in}
\psfig{file=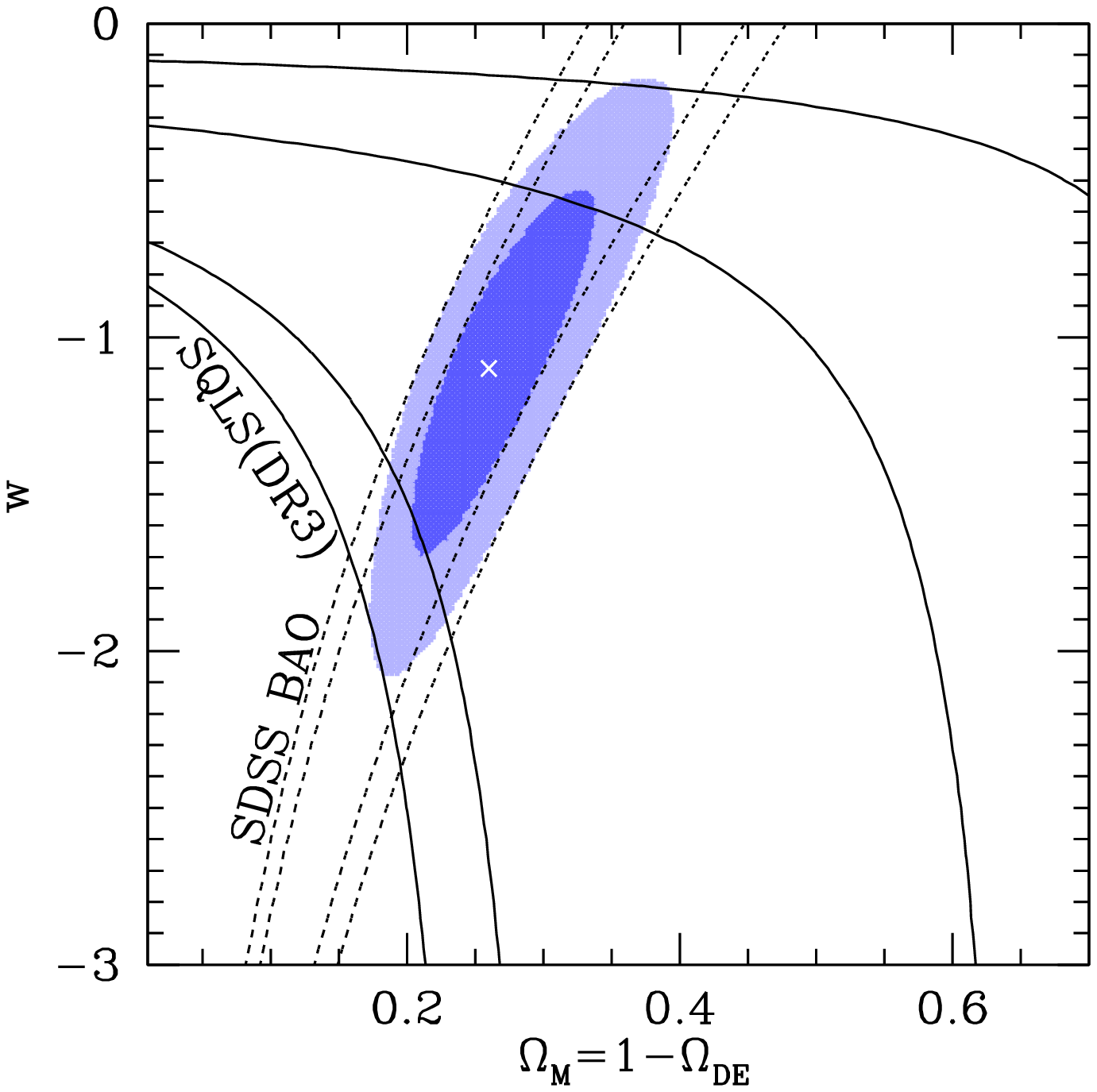,height=2.8in}}
  \caption{Constraints on dark energy from the statistics of strongly 
lensed QSOs. Left panel: constraints on $\Omega_\Lambda, \Omega_m$ 
from the CLASS lens survey, from \citet{mitchell05}. Right panel: constraints 
on $w, \Omega_m$ from the SDSS lensed QSO survey (black), from SDSS BAO (dotted), 
and the combination of the two (blue), from \citet{oguri07}.
}
\label{fig:lens}
\end{figure}

{\bf Strong Lensing:} A distant quasar lying near the line of sight to a 
foreground galaxy can have its light strongly bent by the galaxy's gravitational field.
In favorable 
circumstances, this leads to the appearance of multiple images of the 
same quasar, an instance of strong gravitational lensing.  
Schematically, the optical depth for lensing can be written 

\begin{equation}
\tau(z_S) = \int dV \int dM {dn \over dM} A_L(M) ~,
 \label{eq:lens}
\end{equation}

\noindent where $dV$ is the volume element, 
the volume integral is taken out to the QSO redshift $z_S$, $dn/dM$ is the 
redshift-dependent mass function of the lens population 
(assumed to be massive galaxies and 
their associated dark matter halos), and $A_L(M)$ is determined by the 
density profiles of the lenses, often modeled as singular isothermal 
spheres for galaxy-scale lenses. In principle, the volume element is 
strongly sensitive to the cosmological constant (see Fig. \ref{fig:rz}), 
so the optical depth 
for strong lensing would seem a natural dark energy probe \citep{fukugita92,Kochanek_96}. 
However, 
a major difficulty arises from the fact that the mass function of the 
lens population is not an observable. Traditionally, galaxy scaling relations 
such as the Faber-Jackson relation between luminosity and velocity dispersion, 
along with measurements of the galaxy 
luminosity function $dn/dL$, were used to infer the galaxy mass function $dn/dM$, but the 
associated uncertainties, particularly for galaxies at $z \sim 0.5$ 
that dominate the optical depth, were large. More recent approaches 
have made use of the velocity-dispersion distribution 
function of early-type galaxies, $dn/d\sigma_v$, measured, e.g., by 
the SDSS \citep{mitchell05,oguri07,chae07}. Fig. \ref{fig:lens} shows examples 
of recent constraints on $\Omega_\Lambda$ and $\Omega_m$ (for $w=-1$) and 
on $\Omega_m$ and $w$ (for a flat Universe) from strong lens statistics. 
While the statistical errors are large, the results are consistent 
with an accelerating Universe.

\section{Models of Cosmic Acceleration}

Understanding the origin of cosmic acceleration presents both a challenge 
and an opportunity to theorists. While there has been no shortage of 
ideas, there is also no consensus model. In this section, we briefly 
review the theoretical landscape.

\subsection{Dark Energy Models}

\subsubsection{Vacuum Energy}

Vacuum energy is the simplest candidate for dark energy. 
As noted above, since the stress-energy 
of the vacuum $T_{\mu \nu}^{vac}$ is proportional to the metric $g_{\mu \nu}$, 
it is mathematically equivalent to a cosmological constant. The difficulty 
arises when one attempts to calculate its expected value. For each mode of a quantum field 
there is a zero-point energy $\hbar \omega
/2$, so that the energy density of the quantum vacuum is given by 
\begin{equation}
\rho_{vac} = {1\over 2}\sum_{\rm fields}g_i\int_0^\infty 
\sqrt{k^2 + m^2}\,{d^3k\over (2\pi )^3} \simeq  \sum_{\rm fields} {g_i k_{\rm max}^4\over 16\pi^2}~,
\end{equation}
\noindent where $g_i$ accounts for the degrees of freedom of the field (the sign of
$g_i$ is $+$ for bosons and $-$ for fermions), and the sum runs over all
quantum fields (quarks, leptons, gauge fields, etc). Here $k_{\rm max}$ is an
imposed momentum cutoff, because the sum diverges quartically.

To illustrate the magnitude of the problem, if the energy density contributed
by just one field is to be at most the critical density, then the cutoff
$k_{\rm max}$ must be $< 0.01\,{\rm eV}$ --- well below the energy scale where
one could have appealed to ignorance of physics beyond. Taking the cutoff to be 
the Planck scale
($\approx 10^{19}$\,GeV), where one expects quantum field theory in a classical
spacetime metric to break down, the zero-point energy density would exceed the
critical density by 120 orders of magnitude. It is highly unlikely that 
a classical contribution to the vacuum energy density would cancel this 
quantum contribution to such high precision. This large discrepancy
is known as the cosmological constant problem \citep{WeinbergRMP}. 

Supersymmetry, the hypothetical symmetry between bosons and fermions, appears 
to provide only partial help.
In a supersymmetric (SUSY) world, every fermion in the standard
model of particle physics has an equal-mass SUSY bosonic partner and vice
versa, so that fermionic and bosonic zero-point contributions to $\rho_{vac}$ 
would exactly cancel.  However, SUSY is not a manifest symmetry in
Nature: none of the SUSY particles has yet been observed in collider
experiments, so they must be substantially heavier than their standard-model
partners. If SUSY is spontaneously broken at a mass scale $M$, one expects
the imperfect cancellations to generate a finite vacuum energy density
$\rho_{vac} \sim M^4$.  For the currently favored value $M \sim 1$ TeV,
this leads to a discrepancy of 60 (as opposed to 120) orders of magnitude with
observations. 

One approach to the cosmological constant problem involves the idea that
the vacuum energy scale is a random variable that can take on
different values in different disconnected regions of the Universe.  Because a
value much larger than that needed to explain the observed cosmic 
acceleration, $\rho_{vac} > 125 \rho_{crit}$, 
would preclude the formation of galaxies (assuming all other cosmological
parameters are held fixed), we could not find ourselves in a region with such
large $\rho_{vac}$ \citep{Weinberg87}. Imagining an ensemble of 
universes or of such disconnected large regions, the probability for 
us to observe a particular value of $\rho_{vac}$ is given by \citep{Weinberg87}
$dP(\rho_{vac})=P^*(\rho_{vac})N(\rho_{vac})d \rho_{vac}$, where 
$P^*(\rho_{vac})$ is the prior probability for a region to 
have a given value of 
the vacuum energy density. Here, 
$N(\rho_{vac})$ is the fraction of baryons that 
end up in galaxies or in systems large enough to sustain observers
and has support only at $\rho_{vac}<125 \rho_{crit}$ (again holding 
all other parameters fixed, including the amplitude of primordial 
perturbations). Weinberg \citep{Weinberg87} assumed that $P^*$ is 
broad and 
effectively constant over the anthropically allowed range of $N$. 
Vilenkin and Rubakov, however, have noted that $P^*$ could vary 
strongly over this range and could in fact be strongly peaked at large 
values of $\rho_{vac}$. In that case, we would be more likely 
to find ourselves living ``on the edge'' of the allowed region, 
and life should be nasty, brutish, and short, a view one might 
term the misanthropic principle.  

The anthropic approach finds a
possible home in the landscape version of string theory, in which the number of
different vacuum states is very large and essentially all values of the
cosmological constant are possible.  Provided that the Universe has such a
multiverse structure, this might provide an explanation for the smallness of
the cosmological constant \citep{Bousso_Polchinski,Susskind}. 

\subsubsection{Light Scalar Fields}

Another approach to dark energy involves the idea that the Universe is not yet 
in its ground state. Suppose the true vacuum energy is zero (for reasons yet 
unknown), $\Lambda=0$. Transient vacuum-like energy can exist if there is a 
field that takes a cosmologically long time to reach its ground 
state \citep{Ratra_Peebles,Wetterich,Frieman_PNGB,Zlatev}. This 
was the reasoning behind primordial inflation, a proposed epoch of accelerated 
expansion in the very early universe. For this reasoning to apply now, we must 
postulate the existence of an extremely light scalar field $\phi$, 
since the dynamical 
timescale for evolution of such a field is given by $t_\phi \sim 1/m_\phi$. To 
satisfy $t_\phi > 1/H_0$, the scalar mass should satisfy 
$m_\phi < H_0 \sim 10^{-33}$ eV, extremely tiny by particle physics 
standards for fields that are not exactly massless due to a symmetry.
Since the Compton wavelength of the field is also of 
order $1/H_0 = 3000 h^{-1}$ Mpc or larger, it will not gravitationally 
cluster with large-scale structure---we expect it to be nearly 
smoothly distributed---though it can have small-amplitude perturbations 
on the largest observable scales today, which can affect the 
CMB anisotropy, e.g., \citep{Coble}.

For a
scalar field $\phi$, with Lagrangian density ${\cal L} = {1\over
2}\partial^\mu\phi\partial_\mu\phi - V(\phi )$, the stress-energy takes the
form of a perfect fluid, with
\begin{equation}
\rho = \dot{\phi}^2/2  + V(\phi ) ~~,~~~~ p    = \dot{\phi}^2/2  - V(\phi )~,
\label{field_pot_kin}
\end{equation}
\noindent where $\phi$ is assumed to be spatially homogeneous, i.e., $\phi({\vec
x},t)=\phi(t)$, $\dot\phi^2/2$ is the kinetic energy, and $V(\phi )$ is the
potential energy. The evolution of the field is
governed by its equation of motion,
\begin{equation}
  \ddot {\phi} + 3 H\dot{\phi} + V'(\phi) = 0~, 
\label{eq:field_roll}
\end{equation}
\noindent where a prime denotes differentiation with respect to $\phi$. 
Scalar-field dark energy can be described by the equation-of-state parameter
\begin{equation}
w_\phi = {\dot\phi^2/2 - V(\phi ) \over \dot\phi^2/2 + V(\phi )} =
{-1 + \dot\phi^2/2V \over 1 + \dot\phi^2/2V}~.
\label{eq:wphi}
\end{equation}
\noindent If the scalar field evolves slowly, $\dot{\phi}^2/2V \ll 1$, 
as it generally will do when $m_\phi = \sqrt{V''(\phi)} \ll H(t)$, 
then $w_\phi \approx
-1$, and the scalar field behaves like a slowly varying vacuum energy, with
$\rho_{vac}(t) \simeq V[\phi (t)]$. If this inequality is only marginally 
satisfied, however, then the equation of state parameter can deviate 
significantly from $-1$, and it generally evolves in time.

The simplest such model would involve just a free, massive 
scalar field $V(\phi)= m_\phi^2 \phi^2/2$. In this case, in order for 
the field to both supply negative pressure (and therefore drive 
accelerated expansion) and have the correct magnitude of the 
energy density, $\rho \sim 10^{-10}$ eV$^4$, the field amplitude 
must be very large, $\phi \sim 10^{28}$ eV $\sim M_{Pl}$, 
comparable to the Planck mass. This implies that the 
scalar potential is remarkably flat; one measure of this is 
that $m_\phi/\phi \sim 10^{-61}$ or smaller. Moreover, in order 
not to destroy the required flatness of the potential, the  
quartic self-coupling of the field, $\lambda \phi^4/4$, is 
constrained to be extremely small, $\lambda < 10^{-122}$. 
These are generic features of scalar field dark energy models.
Understanding such very small numbers and ratios makes
it challenging to connect scalar field dark energy with particle physics
models. In constructing theories that go beyond the 
standard model of particle
physics, including those that incorporate primordial inflation, model-builders
have been strongly guided by the requirement that any small dimensionless
numbers in the theory should be protected by symmetries from large quantum
corrections; such small numbers are then said to be ``technically 
natural''. Thus far, this kind of model-building
discipline has not been the rule among cosmologists working on dark energy
models.  

One scenario that does attempt to incorporate the naturalness criterion 
has a pseudo-Nambu-Goldstone boson as the dark energy scalar 
\citep{Frieman_PNGB}. In the 
simplest incarnation, a global $U(1)$ symmetry is spontaneously broken 
at a very high energy scale, $f \sim M_{Pl}$, giving rise to a 
massless Nambu-Goldstone boson. If the symmetry is explicitly broken
at a much lower scale $M \sim 10^{-3}$ eV (which is technically 
natural), then the field gets a tiny mass from the explicit breaking, 
with a periodic potential of the form

\begin{equation}
V(\phi) = M^4 \left[1+ \cos \left({\phi \over f}\right) \right]~.
\label{eq:PNGB}
\end{equation}

\noindent Such a field would be a much lighter cousin of the QCD 
axion. Examples of particle physics model-building incorporating 
this idea are given in \citep{Nomura,Hall}.

In the examples above, at early times the field is frozen to its 
initial value by the friction term $3H \dot{\phi}$ in Eqn. \ref{eq:field_roll}, and it acts 
as vacuum energy; when the expansion rate drops 
below $H^2 = V''(\phi)$, the field begins to roll and $w$ evolves 
away from $-1$. In other models, the field instead may 
roll more slowly as time progresses, i.e., the slope of the potential drops 
more rapidly than the Hubble friction term.   
This can happen if, e.g., $V(\phi)$ falls off exponentially or 
as an inverse power-law 
at large $\phi$. These ``thawing'' and ``freezing'' models 
tend to carve out different trajectories of $w(z)$, so that 
precise cosmological measurements might be able to discriminate 
between them \citep{Caldwell_Linder}. 

As Fig.~\ref{fig:scale} shows, through most of the history of the Universe,
dark matter or radiation dominated dark energy by many orders of magnitude.
We happen to live around the time that dark energy has become important.  Is
this coincidence between $\rho_{DE}$ and $\rho_{m}$ an important clue to
understanding cosmic acceleration or just a natural consequence of the
different scalings of cosmic energy densities and the longevity of the
Universe? In some freezing models, the scalar field energy density tracks that
of the dominant component (radiation or matter) at early times and then
dominates at late times, providing a dynamical origin for the coincidence. 
In thawing models, the coincidence is indeed
transitory and just reflects the mass scale of the scalar field.

\subsection{Modified Gravity}

An alternative approach seeks to explain cosmic acceleration not in terms 
of dark energy but as a manifestation of new gravitational physics. 
Instead of adding a new component $T_{\mu \nu}^{DE}$ to the right  
side of the Einstein equations (Eqn. \ref{eq:Ein}), one instead modifies the 
geometric side---schematically, $G_{\mu \nu} \rightarrow G_{\mu \nu} + f(g_{\mu \nu})$.  
A number of
ideas have been explored along these lines, from models motivated by
higher-dimensional theories and string theory
\citep{DGP,Deffayet}
to phenomenological modifications of the
Einstein-Hilbert action of General Relativity \citep{CDTT,Song_Hu_fR}.

As an example, consider the model of \citet{DGP}, which arises by 
assuming that we live in a 3-dimensional brane in a $4+1$-dimensional 
Universe. The action can be written as

\begin{equation}
S=M_5^3 \int d^5X \sqrt{|{\rm det}g_5|}R_5 + M_{Pl}^2 \int d^4 x \sqrt{|
{\rm det}g_4|}R_4 + \int d^4 x \sqrt{|{\rm det}g_4|} L_m~,
\label{eq:DGP}
\end{equation}

\noindent where the first term is the 5-dimensional Einstein-Hilbert 
action, the second describes the curvature of the brane, and the 
third describes the particles of the Standard Model, confined to the 
brane. At large distances, gravity can leak off the 3-brane into the 
bulk, infinite 5th dimension. The cross-over from effective 4D to 5D 
gravity occurs at a scale $r_c = M^2_{Pl}/M_5^3$ and gives rise 
to a modified Friedmann equation,

\begin{equation}
  H^2 \pm {H\over r_c}= {8\pi G \rho \over 3} ~.
\label{eq:DGP2}
\end{equation}

\noindent Choosing the minus sign in the second term, which becomes 
important when $H \sim r_c^{-1}$, one finds an asymptotically 
self-accelerating solution, $H \rightarrow H_\infty = r_c^{-1}$, even 
though there is no cosmological constant or vacuum energy term 
in the action. For acceleration to set in at recent epochs 
requires the five-dimensional gravitational scale to be 
of order $M_5 \sim 1$ GeV. While attractive, it is not clear that a 
consistent model with 
this dynamical behavior exists \citep{Gregory07}.

In the phenomenological approach, one modifies the Einstein-Hilbert 
action,

\begin{equation}
S=\int d^4 x \sqrt{|{\rm det}g|}R \rightarrow \int d^4 x \sqrt{|{\rm det}g|}
\left[M_{Pl}^2R 
+ f(R,R_{\mu \nu},R_{\mu \nu \alpha \beta})\right]~,
\label{eq:fR}
\end{equation}

\noindent and looks for suitable choices of $f(R,...)$ that induce late 
acceleration \citep{CDTT,Song_Hu_fR}. A challenge for this approach 
is that the physical effects which typically give rise to corrections 
to the gravitational action, e.g., the effects of quantum fields in 
curved spacetime, generally involve positive quadratic forms, e.g., 
$f(R) \sim R^2 + ...$. In that case, by dimensional analysis, the 
correction is only important at very high values of the curvature, 
$R \sim M_{Pl}^2$. To obtain effects at very low curvature, i.e., 
late cosmic epochs, requires inverse powers of the curvature 
invariants entering in $f$. Solar system tests of General Relativity 
place stringent constraints on such models. 

An interesting feature of modified gravity theories is that they 
typically imply a modification of the General Relativity relationship between the 
growth rate of large-scale density perturbations, $\delta \equiv \delta \rho_m({\bf x},t)/\bar{\rho}_m$,
and the cosmic expansion rate $H(t)$. In General Relativity, 
the growth of small-amplitude, matter-density perturbations on 
scales smaller than the Hubble radius is governed by 

\begin{equation}
\ddot{\delta} + 2 H \dot{\delta}-4\pi G \bar{\rho}_m \delta = 0 ~.
\label{eq:growth}
\end{equation}

\noindent In the context of General Relativity, dark energy affects the 
growth of structure through its impact on the expansion rate $H$. If acceleration 
instead arises from a modification of General Relativity, those modifications 
lead to additional terms in Eqn. \ref{eq:growth} that can directly affect 
the growth rate of large-scale structure. As a result, comparing probes 
of the expansion rate, e.g., through cosmic distance measurements, with 
probes of the growth rate of large-scale structure can in principle test 
the consistency of General Relativity (plus dark energy) as the explanation 
for acceleration.

\subsection{Anthropocentric Universe}

Instead of modifying the right or left side of the Einstein equations to 
explain the supernova observations, a third logical possibility is to 
drop the assumption that the Universe is spatially homogeneous on large scales. 
It has been argued that the non-linear gravitational effects of spatial 
density perturbations, when averaged over large scales, could yield 
a distance-redshift relation in our observable patch of the Universe 
that is very similar to that for an accelerating, homogeneous 
Universe \citep{Kolb_Matarrese_06}, obviating the need for either dark energy 
or modified gravity. While there has been 
debate about the amplitude of these effects, this idea has helped 
spark renewed interest in a class of exact, inhomogeneous cosmologies. 
For such Lema\^{\i}tre-Tolman-Bondi models
to be consistent with the SN data and
not conflict with the isotropy of the CMB, the Milky Way must be near the
center of a very large-scale, nearly spherical, underdense region
\citep{Tomita_00,Alnes_05,Enqvist_07}.  Requiring our galaxy 
to occupy a privileged location, in violation of the spirit of the Copernican 
principle, is not yet theoretically well-motivated. However, it remains 
an interesting empirical question whether such models 
can be made consistent with the wealth of precision
cosmological data \citep{CaldSteb}. 

\section{Probing Dark Energy and Cosmic Acceleration}

Although the phenomenon of accelerated expansion is now well established, 
the underlying physical cause remains a mystery. Is it 
dark energy or modified gravity? The primary 
question we would like to address in the near term is whether the 
cosmological constant (or vacuum energy) can be excluded as the 
explanation of acceleration: is General Relativity plus dark energy 
with $w=-1$ viable or not? Such a model is consistent with all the 
extant data, so the possibility of excluding it will require much more 
precise measurements of both the history of the cosmic expansion rate 
and the history of the growth of large-scale structure. To illustrate 
the challenge, consider that 
for fixed $\Omega_{DE}$, a 1\% change in (constant)
$w$ translates to only a 3\% (0.3\%) change in dark-energy (total) density at redshift 
$z=2$ and only a 0.2\% change in distances to redshifts $z=1-2$.

Four methods hold particular
promise in probing cosmic acceleration: type Ia supernovae,
baryon acoustic oscillations, clusters of galaxies, and weak gravitational
lensing. 
We have described SNe and BAO above; they both provide geometric 
probes of the expansion rate. Clusters and 
weak lensing, which we discuss below, are sensitive to both the expansion rate and 
the growth of structure. As a result, these four probes are 
complementary in terms of both dark energy constraints
as well as the systematic errors to which they are susceptible.  Because of this, 
a multi-pronged approach will be most effective. The goals 
of the next generation of dark energy experiments will 
be to constrain the dark energy equation of state parameter $w$ at the 
few percent level in order to address the questions above.

\subsection{Clusters}

Galaxy clusters are the largest virialized objects in the Universe.
Within the context of the cold dark matter paradigm for the 
formation of large-scale structure, the number density of
cluster-sized dark matter halos as a function of redshift and halo mass can be
accurately predicted from N-body simulations \citep{warren_06,tinker08}. Comparing
these predictions to
large-area cluster surveys that extend to high redshift ($z \sim 1$) can
provide precise constraints on the cosmic expansion history
\citep{wang_98,haiman_01}. 

The redshift distribution of clusters in a survey that selects clusters
according to some observable $O$ with redshift-dependent selection function
$f(O,z)$ is given by
\begin{equation}
\frac{d^{2}N(z)}{dzd\Omega} = \frac{r^2(z)}{H(z)} 
\int^{\infty}_{0}f(O,z)dO\int^{\infty}_{0}p(O|M,z)\frac{dn(z)}{dM}dM ~,
\label{eq:clustercount}
\end{equation}
\noindent where $dn(z)/dM$ is the space density of dark halos in comoving coordinates,
and $p(O|M,z)$ is the mass-observable relation, the probability that a
halo of mass $M$ at redshift $z$ is observed as a cluster with observable
property $O$. The utility of this  probe hinges on the ability 
to robustly associate cluster observables such as X-ray luminosity or
temperature, cluster galaxy richness, Sunyaev-Zel'dovich effect flux decrement,
or weak lensing shear, with cluster mass
\citep[e.g.,][]{borgani_06}. 

The sensitivity of cluster counts to dark energy arises from two factors: {\it
geometry}, the term multiplying the integral in Eqn.~(\ref{eq:clustercount})
is the comoving volume element; and {\it growth of structure}, $dn(z)/dM$
depends on the evolution of density perturbations, cf. Eqn.~\ref{eq:growth}. 
The cluster mass function is also
determined by the primordial spectrum of density perturbations; its
near-exponential dependence upon mass is the root of the power of 
clusters to probe dark energy.

\begin{figure}
\centerline{
\psfig{file=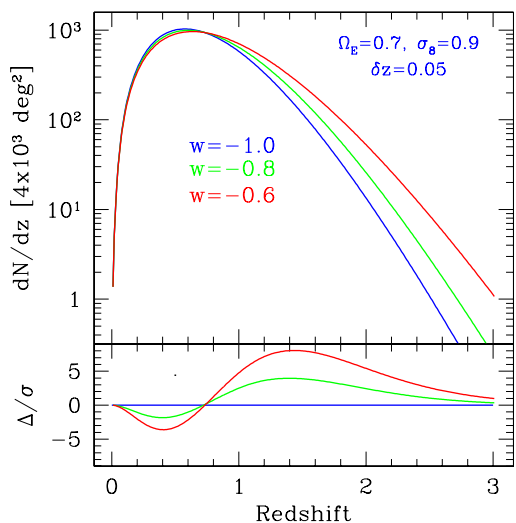,width=3.in}
}
\caption{ 
Predicted cluster counts for a survey covering 4,000 sq.\ deg.\ 
that is sensitive to halos more massive than $2\times 10^{14} M_\odot$, for 3 flat cosmological models with fixed $\Omega_{m}=0.3$ and $\sigma_8=0.9$. 
Lower panel shows differences between the models relative to the 
statistical errors. 
From \citet{Mohr_04}.
}
\label{fig:massfun}
\end{figure}

Fig.~\ref{fig:massfun} shows the sensitivity to the dark energy
equation of state parameter of the expected cluster counts for the South Pole Telescope
and the Dark Energy Survey. At modest redshift, $z<0.6$, the differences are
dominated by the volume element; at higher redshift, the counts are most
sensitive to the growth rate of perturbations.

The primary systematic concerns for the cluster method 
are uncertainties in the mass-observable
relation $p(O|M,z)$ and in the selection function $f(O,z)$.  The strongest
cosmological constraints arise for those cluster 
observables that are most strongly
correlated with mass, i.e., for which $p(O|M,z)$ is narrow for fixed $M$, 
and which have a well-determined selection function.
There are several independent techniques both for detecting clusters and for
estimating their masses using observable proxies.  Future surveys will aim to
combine two or more of these techniques to cross-check cluster mass estimates
and thereby control systematic error. Measurement of the spatial correlations 
of clusters and of the shape of the mass function provide additional 
internal calibration of the mass-observable relation \citep{Majumdar_04,Lima_04}.

\subsection{Weak Lensing}

The gravitational bending of light by structures in the Universe distorts or
shears the images of distant galaxies. 
This distortion allows the distribution of dark matter and its evolution 
with time to be measured, thereby probing the influence of dark energy on 
the growth of structure. The
statistical signal due to gravitational lensing by large-scale structure is
termed ``cosmic shear.''  The cosmic shear field at a point in the sky is
estimated by locally averaging the shapes of large numbers of distant galaxies.
The primary statistical measure of the cosmic shear is the shear angular power
spectrum measured as a function of source-galaxy redshift $z_s$.  (Additional
information is obtained by measuring the correlations between shears at
different redshifts or with foreground lensing galaxies.)  The shear angular
power spectrum is \citep{Kaiser92,hujain03}
\begin{equation}
P^\gamma_{\ell}(z_s) = \int_{0}^{z_s} dz  {H(z) \over d_A^2(z)} |W(z,z_s)|^2
P_\rho\left (k={\ell \over d_A(z)}; z\right )\,,
\label{eqn:Limber}
\end{equation}

\noindent where $\ell$ denotes the angular multipole,
 the weight function $W(z,z_s)$ is the efficiency 
for lensing a population of source galaxies and is determined by the distance 
distributions 
of the source and lens galaxies, 
and $P_\rho(k,z)$ is the power spectrum of density
perturbations.  

\begin{figure}
\centerline{\psfig{file=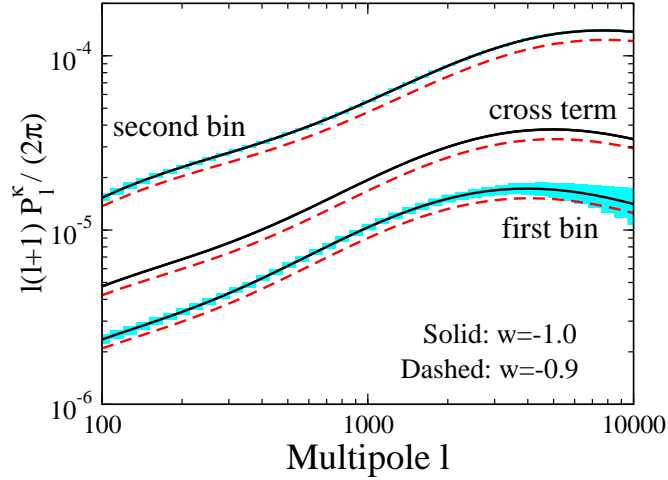,width=3.6in,height=2.8in,angle=-90}}
\caption{
Cosmic shear angular power spectrum and statistical errors 
expected for LSST 
for $w=-1$ and $-0.9$. For 
illustration, 
results are shown for source galaxies in two broad redshift bins,  
$z_s=0-1$ (first bin) and $z_s=1-3$ (second bin); the cross-power 
spectrum between the two bins (cross term) is shown without the 
statistical errors. From \citet{fth08}.
}
\label{fig:P_kappa_tomo}
\end{figure}

As with clusters, the dark-energy sensitivity of the shear angular power
spectrum comes from two factors: {\it geometry}---the Hubble parameter, the
angular-diameter distance, and the weight functions; and {\it growth of
structure}---through the evolution of the power spectrum of density
perturbations.  It is also possible to separate these effects and extract a
purely geometric probe of dark energy from the redshift dependence of
galaxy-shear correlations \citep{JainTaylor,Bernstein_Jain}.
The three-point correlation of cosmic shear is also sensitive to 
dark energy \citep{Takada_Jain}.

The statistical uncertainty in measuring the shear power spectrum on
large scales is \citep{Kaiser92}
\begin{equation}
\Delta P^\gamma_\ell = \sqrt{\frac{2}{(2\ell+1)f_{\rm sky}} }
\left[ P^\gamma_\ell +\frac{\sigma^2(\gamma_i)}{n_{\rm eff}} \right]~~,
\label{eqn:power_error}
\end{equation}
where $f_{\rm sky}$ is the fraction of sky area covered by the survey,
$\sigma^2(\gamma_i)$ is the variance in a single component of the
(two-component) shear, and $n_{\rm eff}$ is the effective number density per
steradian of galaxies with well-measured shapes. The first term in brackets,
which dominates on large scales, comes from cosmic variance of the mass
distribution, and the second, shot-noise term results from both the variance
in galaxy ellipticities (``shape noise'') and from shape-measurement errors
due to noise in the images. Fig.~\ref{fig:P_kappa_tomo} shows the dependence
on the dark energy of the shear power spectrum and an indication of the 
statistical errors expected for a survey such as that planned for LSST, assuming  
a survey area of 15,000 sq. deg. and effective source galaxy 
density of 
$n_{\rm eff}=30$ galaxies per sq. arcmin.

Systematic errors in weak lensing measurements arise from a number of sources
\citep{Huterer06}: incorrect shear estimates, uncertainties in galaxy
photometric redshift estimates, intrinsic correlations of galaxy shapes, and
theoretical uncertainties in the mass power spectrum on small scales.  The
dominant cause of galaxy shape measurement error in current lensing surveys is
the anisotropy of the image point spread function (PSF) caused by optical and
CCD distortions, tracking errors, wind shake, atmospheric refraction, etc. 
This error can be diagnosed since there are geometric 
constraints on the shear patterns that can be produced by lensing that 
are not respected by systematic effects. 
A second kind of shear measurement error arises from miscalibration of the
relation between measured galaxy shape and inferred shear,
arising from inaccurate correction for the
circular blurring of galaxy images due to atmospheric seeing. 
Photometric redshift errors impact shear power spectrum estimates primarily
through uncertainties in the scatter and bias of photometric redshift estimates
in redshift bins \citep{Huterer06,Ma06}.  
Any tendency of galaxies to align with their neighbors --- or to align with
the local mass distribution --- can be confused with alignments caused by
gravitational lensing, thus biasing dark energy determinations
\citep{Hirata04,Heymans}.  Finally, uncertainties in the theoretical mass power
spectrum on small scales could complicate attempts to use the high-multipole
($\ell >$ several hundred) shear power spectrum to constrain dark
energy. Fortunately, weak lensing surveys should be able to internally
constrain the impact of such effects \citep{zentner07}.

\section{Dark Energy Projects}

A diverse and ambitious set of projects to probe dark energy are in
progress or being planned. Here we provide a brief overview 
of the observational landscape. Table \ref{tab:surveys}  
provides a representative sampling, not a comprehensive listing, of projects that are 
currently proposed or under construction and does not 
include experiments that have already 
reported results. All of these projects share the common feature of 
surveying wide
areas to collect large samples of objects --- galaxies, clusters, or
supernovae. 

The Dark Energy Task Force (DETF) report \citep{DETF} classified dark energy
surveys into an approximate sequence: on-going projects, either taking data or
soon to be taking data, are Stage II; near-future, intermediate-scale projects
are Stage III; and larger-scale, longer-term future projects are designated
Stage IV. More advanced stages are in general expected to deliver tighter dark
energy constraints. 

\subsection{Ground-based surveys}

A number of projects are underway to detect clusters and probe dark energy using the
Sunyaev-Zel'dovich effect. These surveys are
coordinated with optical surveys that can determine cluster redshifts. The
Atacama Pathfinder EXperiment (APEX) survey in Chile 
will cover up to 1000 square degrees. The largest of these projects
are the Atacama Cosmology Telescope (ACT) and the South Pole Telescope
(SPT), the latter of which will carry out a 4,000 square degree survey.

A number of optical imaging surveys are planned or proposed which can study
dark energy through weak lensing, clusters, and angular BAO using a single
wide-area survey.  These projects use telescopes of intermediate to large
aperture and wide field-of-view, gigapixel-scale CCD cameras, and are deployed
at the best astronomical sites in order to obtain deep galaxy photometry and
shape measurements. They deliver photometric-redshift information through color
measurements using multiple passbands. The ESO VLT Survey Telescope (VST) on
Cerro Paranal will carry out public surveys, including the 1500 sq. deg. KIDS
survey and a shallower, 5000 sq. deg. survey (ATLAS). The Panoramic Survey
Telescope and Rapid Response System (Pan-STARRS)-1 uses a 1.8-m wide-field
telescope to carry out several wide-area surveys from Haleakala; in the future,
they hope to deploy $4 \times 1.8$-m telescopes at Mauna Kea in
Pan-STARRS-4. The Dark Energy Survey (DES) will use a new 3 sq. deg.  imager
with red-sensitive CCDs on a 4-m telescope 
at Cerro Tololo Inter-American Observatory (CTIO) in Chile to
carry out a 5,000 sq. deg. survey in 5 optical passbands, covering the same
survey area as the SPT and partnering with the ESO VISTA Hemisphere Survey
which will survey the same area in 3 near-infrared bands.  Hyper Suprime-Cam is
a new wide-field imager planned for the Subaru telescope on Mauna Kea that will
be used to carry out a deep survey over 2000 sq. deg.  The Advanced
Liquid-mirror Probe of Asteroids, Cosmology and Astrophysics (ALPACA) is a
proposed rotating liquid mercury telescope that would repeatedly survey a long,
narrow strip of the sky at CTIO.  The most ambitious of these projects is the
Large Synoptic Survey Telescope (LSST), which would deploy a multi-Gigapixel
camera with 10 sq. deg.  field-of-view on a new telescope on Cerro Pachon in
Chile to survey 15,000 sq.  deg. over 10 years.

\begin{table}
    \caption{Dark energy projects proposed or under construction. Stage refers 
to the DETF time-scale classification.\vspace{0.2cm}}
    \label{tab:surveys}
    \begin{tabular}{llll} \hline\hline
\rule[-3mm]{0mm}{8mm} Survey & Description & Probes & Stage \\ \hline
Ground-based: && \\
ACT        & SZE, 6-m                                  & CL & II  \\
APEX       & SZE, 12-m                                 & CL & II  \\
SPT        & SZE, 10-m                                 & CL & II  \\
VST        & Optical imaging, 2.6-m                   & BAO,CL,WL & II \\
Pan-STARRS 1(4) & Optical imaging, 1.8-m($\times 4$)  & All & II(III) \\
DES        & Optical imaging, 4-m                     & All & III \\
Hyper Suprime-Cam & Optical imaging, 8-m              & WL,CL,BAO & III \\
ALPACA     & Optical imaging, 8-m                     & SN, BAO, CL & III \\
LSST       & Optical imaging, 6.8-m                   & All & IV \\
AAT WiggleZ& Spectroscopy, 4-m                        & BAO & II \\
HETDEX     & Spectroscopy, 9.2-m                      & BAO & III \\
PAU        & Multi-filter imaging, 2-3-m                     & BAO & III \\
SDSS BOSS  & Spectroscopy, 2.5-m                      & BAO & III \\
WFMOS      & Spectroscopy, 8-m                        & BAO & III \\
HSHS       & 21-cm radio telescope                    & BAO & III \\
SKA        & km$^2$ radio telescope                   & BAO, WL & IV \\
\hline
Space-based: && \\
{\em JDEM Candidates}                               &&   \\
\ \ ADEPT      & Spectroscopy            & BAO, SN & IV  \\         
\ \ DESTINY    & Grism spectrophotometry & SN      & IV  \\ 
\ \ SNAP       & Optical+NIR+spectro     & All  & IV  \\ 
{\em Proposed ESA Missions}                            &&\\
\ \ Euclid       & Imaging \& spectroscopy         & WL, BAO, CL      &     \\
\ \ eROSITA    & X-ray                   & CL      &     \\
{\em CMB Space Probe}                                 && \\
\ \ Planck     & SZE                      & CL      &     \\
{\em Beyond Einstein Probe}                           && \\
\ \ Constellation-X & X-ray              & CL      & IV  \\
\hline\hline
    \end{tabular}
\end{table}

Several large spectroscopic surveys have been designed to detect baryon
acoustic oscillations by measuring $\sim 10^5-10^9$ galaxy and QSO redshifts
using large multi-fiber spectrographs.  WiggleZ is using the Anglo-Australian
Telescope to collect spectra of 400,000 galaxies in the redshift range
$0.5<z<1$.  The Baryon Oscillation Sky Survey (BOSS) will use the SDSS
telescope in New Mexico to measure galaxy spectra out to $z \sim 0.6$.  The Hobby
Eberly Telescope Dark energy EXperiment (HETDEX) plans to target Ly-{$\alpha$}
emitters at higher redshift, $2<z <4$. The Wide-Field
Multi-Object Spectrograph (WFMOS), proposed for the Subaru telescope, would
target galaxies at $z<1.3$ and Lyman-break galaxies at $2.5<
z < 3.5$. The Physics of the Accelerating Universe (PAU) is a Spanish 
project to deploy a wide-field camera with a large number of narrow filters 
to measure coarse-grained galaxy
spectra out to $z=0.9$.

Finally, the proposed Square Kilometer Array (SKA), an array of radio antennas
with unprecedented collecting area, would probe dark energy using baryon acoustic 
oscillations and weak lensing of galaxies via measurements of the 21-cm line
signature of neutral hydrogen (HI). The 
Hubble Sphere Hydrogen Survey (HSHS) aims to carry out a 21-cm BAO 
survey on a shorter timescale.

\begin{figure}
\centerline{\psfig{file=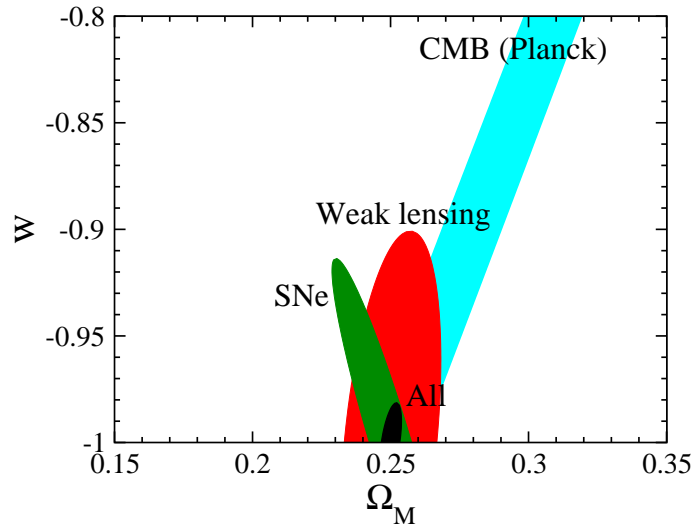,height=2.9in,angle=-90}
}
\caption{Illustration of forecast constraints on dark energy parameters. 
Shown are 68\% C.L.\ uncertainties in the $\Omega_m$ vs. $w$ plane, 
for one version of the proposed SNAP
experiment, which combines a narrow-area survey of 2000 SNe to $z=1.7$ and a
weak lensing survey of 1000 sq. deg.  
From \citet{fth08}. 
}
\label{fig:Omw_and_w0wa}
\end{figure}

\subsection{Space-based surveys}

Three of the proposed space projects are candidates for the Joint Dark Energy
Mission (JDEM), a joint mission of the U.S. Department of Energy (DOE) and the
NASA Beyond Einstein program, targeted at dark energy science.
SuperNova/Acceleration Probe (SNAP) proposes to study dark energy using a
dedicated 2-m class telescope. With imaging in 9 optical and near-infrared passbands
and follow-up spectroscopy of supernovae, it is principally designed to probe
SNe Ia and weak lensing, taking advantage of the excellent optical image
quality and near-infrared transparency of a space-based platform.
Fig.~\ref{fig:Omw_and_w0wa} gives an illustration of the statistical
constraints that the proposed SNAP mission could achieve, by combining SN and
weak lensing observations with results from the Planck CMB mission.  
The Dark Energy Space Telescope (DESTINY) would use a
similar-size telescope with a near-infrared grism spectrograph to study
supernovae.  The Advanced Dark Energy Physics Telescope (ADEPT) is a
spectroscopic mission with the primary goal of constraining dark energy via
baryon acoustic oscillations at $z\sim 2$ as well as supernovae. Another
proposed mission within the NASA Beyond Einstein program is Constellation-X,
which could observe X-ray clusters with unprecedented sensitivity.

There is one European Space Agency (ESA) mission nearing launch and two
concepts under study. The Planck mission, planned for launch in early 2009, 
in addition to pinning down other cosmological parameters important for 
dark energy, will detect thousands of
galaxy clusters using the SZE.  Dark Universe Explorer
(DUNE) and SPACE are optical missions to study dark energy using weak lensing
and baryon acoustic oscillations, respectively, that have 
recently been combined into a single concept mission known 
as Euclid. Finally, the extended ROentgen
Survey with an Imaging Telescope Array (eROSITA), a German-Russian
collaboration, is a planned X-ray telescope that will study dark energy using
the abundance of X-ray clusters.

\section{Conclusion}

The case for an accelerating Universe, which began with the supernova discoveries 
ten years ago, has strengthened into a compelling web of evidence in the years since. 
The simplest explanation for acceleration is dark energy, and the simplest 
candidate for dark energy is vacuum energy---the cosmological constant. 
However, given the lack of understanding of the cosmological constant 
problem, the relative dearth of well-motivated models, and the fact that the Universe 
likely underwent a previous epoch of accelerated expansion (primordial inflation), 
it is best to keep an open mind and rely on experiment as a guide to 
illuminating the underlying cause. Probing the history of cosmic expansion 
and of the growth of structure offers the best hope of pointing us down 
the correct path. An impressive array of experiments with 
that aim are underway or planned, exploiting four primary, complementary 
techniques of probing cosmic acceleration: 
supernovae, baryon acoustic oscillations, clusters, and 
weak lensing. Exploiting the complementarity of these multiple probes 
will be key, since we do not know what the ultimate systematic error floors 
for each method will be. Ten to fifteen years from now, we should know whether 
the effective dark energy equation of state parameter is consistent 
with vacuum energy, that is, $w=-1$, to within a few percent, or not.
The Chinese origin of the phrase ``may you live in interesting times'' 
is apparently unsubstantiated. Let us hope that also calls into doubt 
the third part of the same proverb, ``may you find what you are looking for'', 
since the alternative would certainly be more interesting.


\begin{theacknowledgments}
I would like to thank the organizers of the course, especially 
Jailson Alcaniz, Paulo Pellegrini, and Eduardo Telles, along with 
all the participants, for 
an enjoyable week and a stimulating atmosphere at the Observatorio in Rio. 
I thank Michael Turner and Dragan Huterer for collaborative effort 
on our dark energy review, which helped in the preparation of the lectures 
and of these lecture notes. Research supported by the U.S. Department of Energy at 
Fermilab and at the University of Chicago and by the Kavli Institute for 
Cosmological Physics at Chicago, an NSF Physics Frontier Center.  
\end{theacknowledgments}



\bibliographystyle{aipproc}   

\bibliography{frieman-rio}

\begin{thebibliography}{117}
\expandafter\ifx\csname natexlab\endcsname\relax\def\natexlab#1{#1}\fi
\providecommand{\enquote}[1]{``#1''}
\expandafter\ifx\csname url\endcsname\relax
  \def\url#1{\texttt{#1}}\fi
\expandafter\ifx\csname urlprefix\endcsname\relax\def\urlprefix{URL }\fi
\providecommand{\eprint}[2][]{\url{#2}}

\bibitem[{Riess} et~al.(1998)]{riess98}
A.~G. {Riess}, A.~V. {Filippenko}, P.~{Challis}, A.~{Clocchiatti},
  A.~{Diercks}, P.~M. {Garnavich}, R.~L. {Gilliland}, C.~J. {Hogan}, S.~{Jha},
  R.~P. {Kirshner}, B.~{Leibundgut}, M.~M. {Phillips}, D.~{Reiss}, B.~P.
  {Schmidt}, R.~A. {Schommer}, R.~C. {Smith}, J.~{Spyromilio}, C.~{Stubbs},
  N.~B. {Suntzeff}, and J.~{Tonry}, \emph{AJ} \textbf{116}, 1009--1038 (1998),
  \eprint{arXiv:astro-ph/9805201}.

\bibitem[{Perlmutter} et~al.(1999)]{perlmutter99}
S.~{Perlmutter}, G.~{Aldering}, G.~{Goldhaber}, R.~A. {Knop}, P.~{Nugent},
  P.~G. {Castro}, S.~{Deustua}, S.~{Fabbro}, A.~{Goobar}, D.~E. {Groom}, I.~M.
  {Hook}, A.~G. {Kim}, M.~Y. {Kim}, J.~C. {Lee}, N.~J. {Nunes}, R.~{Pain},
  C.~R. {Pennypacker}, R.~{Quimby}, C.~{Lidman}, R.~S. {Ellis}, M.~{Irwin},
  R.~G. {McMahon}, P.~{Ruiz-Lapuente}, N.~{Walton}, B.~{Schaefer}, B.~J.
  {Boyle}, A.~V. {Filippenko}, T.~{Matheson}, A.~S. {Fruchter}, N.~{Panagia},
  H.~J.~M. {Newberg}, W.~J. {Couch}, and {The Supernova Cosmology Project},
  \emph{ApJ} \textbf{517}, 565--586 (1999), \eprint{arXiv:astro-ph/9812133}.

\bibitem[Copeland et~al.(2006)]{Copeland_review}
E.~J. Copeland, M.~Sami, and S.~Tsujikawa, \emph{Int. J. Mod. Phys.}
  \textbf{D15}, 1753--1936 (2006), \eprint{hep-th/0603057}.

\bibitem[Padmanabhan(2003)]{Padmanabhan_review}
T.~Padmanabhan, \emph{Phys. Rept.} \textbf{380}, 235--320 (2003),
  \eprint{hep-th/0212290}.

\bibitem[Peebles and Ratra(2003)]{Peebles_Ratra_03}
P.~J.~E. Peebles, and B.~Ratra, \emph{Rev. Mod. Phys.} \textbf{75}, 559--606
  (2003), \eprint{astro-ph/0207347}.

\bibitem[Uzan(2007)]{Uzan_06}
J.-P. Uzan, \emph{Gen. Rel. Grav.} \textbf{39}, 307--342 (2007),
  \eprint{astro-ph/0605313}.

\bibitem[Huterer and Turner(2001)]{Hut_Tur_00}
D.~Huterer, and M.~S. Turner, \emph{Phys. Rev.} \textbf{D64}, 123527 (2001),
  \eprint{astro-ph/0012510}.

\bibitem[{Sahni} and {Starobinsky}(2006)]{Sahni_review}
V.~{Sahni}, and A.~{Starobinsky}, \emph{astro-ph/0610026}  (2006),
  \eprint{astro-ph/0610026}.

\bibitem[Linder(2007)]{Linder_review}
E.~V. Linder, \emph{astro-ph/0704.2064}  (2007), \eprint{arXiv:0704.2064
  [astro-ph]}.

\bibitem[{Carroll} et~al.(1992)]{carroll92}
S.~M. {Carroll}, W.~H. {Press}, and E.~L. {Turner}, \emph{ARAA} \textbf{30},
  499--542 (1992).

\bibitem[Carroll(2001)]{Carroll_LivRevRel}
S.~M. Carroll, \emph{Living Rev. Rel.} \textbf{4}, 1 (2001),
  \eprint{astro-ph/0004075}.

\bibitem[Weinberg(1989)]{WeinbergRMP}
S.~Weinberg, \emph{Rev. Mod. Phys.} \textbf{61}, 1--23 (1989).

\bibitem[{Frieman} et~al.(2008{\natexlab{a}})]{fth08}
J.~{Frieman}, M.~{Turner}, and D.~{Huterer}, \emph{ArXiv e-prints} \textbf{803}
  (2008{\natexlab{a}}), \eprint{0803.0982}.

\bibitem[{Dodelson}(2003)]{Dodelson_book}
S.~{Dodelson}, \emph{{Modern cosmology}}, Amsterdam (Netherlands): Academic
  Press, 2003.

\bibitem[{Kolb} and {Turner}(1990)]{Kolb_Turner}
E.~W. {Kolb}, and M.~S. {Turner}, \emph{{The early universe}}, Reading, MA:
  Addison-Wesley, 1990.

\bibitem[{Peacock}(1999)]{Peacock_book}
J.~A. {Peacock}, \emph{{Cosmological Physics}}, Cambridge, UK: Cambridge
  University Press, 1999.

\bibitem[{Peebles}(1993)]{Peebles_book}
P.~J.~E. {Peebles}, \emph{{Principles of physical cosmology}}, Princeton, NJ:
  Princeton University Press, 1993.

\bibitem[{Freedman} et~al.(2001)]{Freedmanetal}
W.~L. {Freedman}, B.~F. {Madore}, B.~K. {Gibson}, L.~{Ferrarese}, D.~D.
  {Kelson}, S.~{Sakai}, J.~R. {Mould}, R.~C. {Kennicutt}, Jr., H.~C. {Ford},
  J.~A. {Graham}, J.~P. {Huchra}, S.~M.~G. {Hughes}, G.~D. {Illingworth}, L.~M.
  {Macri}, and P.~B. {Stetson}, \emph{ApJ} \textbf{553}, 47--72 (2001),
  \eprint{arXiv:astro-ph/0012376}.

\bibitem[Zel'dovich(1968)]{zeldovich68}
Y.~B. Zel'dovich, \emph{Sov. Phys. Usp.} \textbf{11}, 381--393 (1968).

\bibitem[Guth(1981)]{guth80}
A.~H. Guth, \emph{Phys. Rev.} \textbf{D23}, 347--356 (1981).

\bibitem[{Efstathiou} et~al.(1990)]{efstathiou90}
G.~{Efstathiou}, W.~J. {Sutherland}, and S.~J. {Maddox}, \emph{Nature}
  \textbf{348}, 705--707 (1990).

\bibitem[Frieman et~al.(1995)]{Frieman_PNGB}
J.~A. Frieman, C.~T. Hill, A.~Stebbins, and I.~Waga, \emph{Phys. Rev. Lett.}
  \textbf{75}, 2077--2080 (1995), \eprint{astro-ph/9505060}.

\bibitem[Krauss and Turner(1995)]{Krauss_Turner}
L.~M. Krauss, and M.~S. Turner, \emph{Gen. Rel. Grav.} \textbf{27}, 1137--1144
  (1995), \eprint{astro-ph/9504003}.

\bibitem[{Ostriker} and {Steinhardt}(1995)]{JPO_PJS}
J.~P. {Ostriker}, and P.~J. {Steinhardt}, \emph{Nature} \textbf{377}, 600--602
  (1995).

\bibitem[{Hillebrandt} and {Niemeyer}(2000)]{Hillebrandt_00}
W.~{Hillebrandt}, and J.~C. {Niemeyer}, \emph{ARAA} \textbf{38}, 191--230
  (2000), \eprint{arXiv:astro-ph/0006305}.

\bibitem[{Arnett}(1982)]{Arnett_82}
W.~D. {Arnett}, \emph{ApJ} \textbf{253}, 785--797 (1982).

\bibitem[{Hoeflich}(2004)]{hoflich04}
P.~{Hoeflich}, \emph{astro-ph/0409170}  (2004).

\bibitem[{Plewa} et~al.(2004)]{plewa}
T.~{Plewa}, A.~C. {Calder}, and D.~Q. {Lamb}, \emph{ApJ} \textbf{612}, L37--L40
  (2004), \eprint{arXiv:astro-ph/0405163}.

\bibitem[{Phillips}(1993)]{Phillips_93}
M.~M. {Phillips}, \emph{ApJ} \textbf{413}, L105--L108 (1993).

\bibitem[{Riess}(2000)]{Riess_2000}
A.~G. {Riess}, \emph{PASP} \textbf{112}, 1284 (2000),
  \eprint{arXiv:astro-ph/0005229}.

\bibitem[{Perlmutter} and {Schmidt}(2003)]{Perlmutter_Schmidt}
S.~{Perlmutter}, and B.~P. {Schmidt}, \enquote{{Measuring Cosmology with
  Supernovae},} in \emph{Supernovae and Gamma-Ray Bursters}, edited by
  K.~{Weiler}, 2003, vol. 598 of \emph{Lecture Notes in Physics, Berlin
  Springer Verlag}, pp. 195--217.

\bibitem[{Dunkley} et~al.(2008)]{dunkley08}
J.~{Dunkley}, E.~{Komatsu}, M.~R. {Nolta}, D.~N. {Spergel}, D.~{Larson},
  G.~{Hinshaw}, L.~{Page}, C.~L. {Bennett}, B.~{Gold}, N.~{Jarosik}, J.~L.
  {Weiland}, M.~{Halpern}, R.~S. {Hill}, A.~{Kogut}, M.~{Limon}, S.~S. {Meyer},
  G.~S. {Tucker}, E.~{Wollack}, and E.~L. {Wright}, \emph{ArXiv e-prints}
  \textbf{803} (2008), \eprint{0803.0586}.

\bibitem[{Jaffe} et~al.(2001)]{Boom}
A.~H. {Jaffe}, P.~A. {Ade}, A.~{Balbi}, J.~J. {Bock}, J.~R. {Bond},
  J.~{Borrill}, A.~{Boscaleri}, K.~{Coble}, B.~P. {Crill}, P.~{de Bernardis},
  P.~{Farese}, P.~G. {Ferreira}, K.~{Ganga}, M.~{Giacometti}, S.~{Hanany},
  E.~{Hivon}, V.~V. {Hristov}, A.~{Iacoangeli}, A.~E. {Lange}, A.~T. {Lee},
  L.~{Martinis}, S.~{Masi}, P.~D. {Mauskopf}, A.~{Melchiorri}, T.~{Montroy},
  C.~B. {Netterfield}, S.~{Oh}, E.~{Pascale}, F.~{Piacentini}, D.~{Pogosyan},
  S.~{Prunet}, B.~{Rabii}, S.~{Rao}, P.~L. {Richards}, G.~{Romeo}, J.~E.
  {Ruhl}, F.~{Scaramuzzi}, D.~{Sforna}, G.~F. {Smoot}, R.~{Stompor}, C.~D.
  {Winant}, and J.~H. {Wu}, \emph{Physical Review Letters} \textbf{86},
  3475--3479 (2001), \eprint{arXiv:astro-ph/0007333}.

\bibitem[{Pryke} et~al.(2002)]{DASI}
C.~{Pryke}, N.~W. {Halverson}, E.~M. {Leitch}, J.~{Kovac}, J.~E. {Carlstrom},
  W.~L. {Holzapfel}, and M.~{Dragovan}, \emph{ApJ} \textbf{568}, 46--51 (2002),
  \eprint{arXiv:astro-ph/0104490}.

\bibitem[{Hanany} et~al.(2000)]{MAXIMA}
S.~{Hanany}, P.~{Ade}, A.~{Balbi}, J.~{Bock}, J.~{Borrill}, A.~{Boscaleri},
  P.~{de Bernardis}, P.~G. {Ferreira}, V.~V. {Hristov}, A.~H. {Jaffe}, A.~E.
  {Lange}, A.~T. {Lee}, P.~D. {Mauskopf}, C.~B. {Netterfield}, S.~{Oh},
  E.~{Pascale}, B.~{Rabii}, P.~L. {Richards}, G.~F. {Smoot}, R.~{Stompor},
  C.~D. {Winant}, and J.~H.~P. {Wu}, \emph{ApJ} \textbf{545}, L5--L9 (2000),
  \eprint{arXiv:astro-ph/0005123}.

\bibitem[{Komatsu} et~al.(2008)]{komatsu08}
E.~{Komatsu}, J.~{Dunkley}, M.~R. {Nolta}, C.~L. {Bennett}, B.~{Gold},
  G.~{Hinshaw}, N.~{Jarosik}, D.~{Larson}, M.~{Limon}, L.~{Page}, D.~N.
  {Spergel}, M.~{Halpern}, R.~S. {Hill}, A.~{Kogut}, S.~S. {Meyer}, G.~S.
  {Tucker}, J.~L. {Weiland}, E.~{Wollack}, and E.~L. {Wright}, \emph{ArXiv
  e-prints} \textbf{803} (2008), \eprint{0803.0547}.

\bibitem[{Eisenstein} et~al.(2005)]{SDSS_BAO}
D.~J. {Eisenstein}, I.~{Zehavi}, D.~W. {Hogg}, R.~{Scoccimarro}, M.~R.
  {Blanton}, R.~C. {Nichol}, R.~{Scranton}, H.-J. {Seo}, M.~{Tegmark},
  Z.~{Zheng}, S.~F. {Anderson}, J.~{Annis}, N.~{Bahcall}, J.~{Brinkmann},
  S.~{Burles}, F.~J. {Castander}, A.~{Connolly}, I.~{Csabai}, M.~{Doi},
  M.~{Fukugita}, J.~A. {Frieman}, K.~{Glazebrook}, J.~E. {Gunn}, J.~S.
  {Hendry}, G.~{Hennessy}, Z.~{Ivezi{\'c}}, S.~{Kent}, G.~R. {Knapp}, H.~{Lin},
  Y.-S. {Loh}, R.~H. {Lupton}, B.~{Margon}, T.~A. {McKay}, A.~{Meiksin}, J.~A.
  {Munn}, A.~{Pope}, M.~W. {Richmond}, D.~{Schlegel}, D.~P. {Schneider},
  K.~{Shimasaku}, C.~{Stoughton}, M.~A. {Strauss}, M.~{SubbaRao}, A.~S.
  {Szalay}, I.~{Szapudi}, D.~L. {Tucker}, B.~{Yanny}, and D.~G. {York},
  \emph{ApJ} \textbf{633}, 560--574 (2005), \eprint{arXiv:astro-ph/0501171}.

\bibitem[{Percival} et~al.(2007)]{percival07}
W.~J. {Percival}, S.~{Cole}, D.~J. {Eisenstein}, R.~C. {Nichol}, J.~A.
  {Peacock}, A.~C. {Pope}, and A.~S. {Szalay}, \emph{MNRAS} \textbf{381},
  1053--1066 (2007), \eprint{arXiv:0705.3323}.

\bibitem[{Tegmark} et~al.(2006)]{SDSS_LRG}
M.~{Tegmark}, D.~J. {Eisenstein}, M.~A. {Strauss}, D.~H. {Weinberg}, M.~R.
  {Blanton}, J.~A. {Frieman}, M.~{Fukugita}, J.~E. {Gunn}, A.~J.~S. {Hamilton},
  G.~R. {Knapp}, R.~C. {Nichol}, J.~P. {Ostriker}, N.~{Padmanabhan}, W.~J.
  {Percival}, D.~J. {Schlegel}, D.~P. {Schneider}, R.~{Scoccimarro},
  U.~{Seljak}, H.-J. {Seo}, M.~{Swanson}, A.~S. {Szalay}, M.~S. {Vogeley},
  J.~{Yoo}, I.~{Zehavi}, K.~{Abazajian}, S.~F. {Anderson}, J.~{Annis}, N.~A.
  {Bahcall}, B.~{Bassett}, A.~{Berlind}, J.~{Brinkmann}, T.~{Budavari},
  F.~{Castander}, A.~{Connolly}, I.~{Csabai}, M.~{Doi}, D.~P. {Finkbeiner},
  B.~{Gillespie}, K.~{Glazebrook}, G.~S. {Hennessy}, D.~W. {Hogg}, {\v
  Z}.~{Ivezi{\'c}}, B.~{Jain}, D.~{Johnston}, S.~{Kent}, D.~Q. {Lamb}, B.~C.
  {Lee}, H.~{Lin}, J.~{Loveday}, R.~H. {Lupton}, J.~A. {Munn}, K.~{Pan},
  C.~{Park}, J.~{Peoples}, J.~R. {Pier}, A.~{Pope}, M.~{Richmond},
  C.~{Rockosi}, R.~{Scranton}, R.~K. {Sheth}, A.~{Stebbins}, C.~{Stoughton},
  I.~{Szapudi}, D.~L. {Tucker}, D.~E.~V. {Berk}, B.~{Yanny}, and D.~G. {York},
  \emph{Phys. Rev. D} \textbf{74}, 123507 (2006),
  \eprint{arXiv:astro-ph/0608632}.

\bibitem[{Kowalski} et~al.(2008)]{kowalski08}
M.~{Kowalski}, D.~{Rubin}, G.~{Aldering}, R.~J. {Agostinho}, A.~{Amadon},
  R.~{Amanullah}, C.~{Balland}, K.~{Barbary}, G.~{Blanc}, P.~J. {Challis},
  A.~{Conley}, N.~V. {Connolly}, R.~{Covarrubias}, K.~S. {Dawson}, S.~E.
  {Deustua}, R.~{Ellis}, S.~{Fabbro}, V.~{Fadeyev}, X.~{Fan}, B.~{Farris},
  G.~{Folatelli}, B.~L. {Frye}, G.~{Garavini}, E.~L. {Gates}, L.~{Germany},
  G.~{Goldhaber}, B.~{Goldman}, A.~{Goobar}, D.~E. {Groom}, J.~{Haissinski},
  D.~{Hardin}, I.~{Hook}, S.~{Kent}, A.~G. {Kim}, R.~A. {Knop}, C.~{Lidman},
  E.~V. {Linder}, J.~{Mendez}, J.~{Meyers}, G.~J. {Miller}, M.~{Moniez}, A.~M.
  {Mourao}, H.~{Newberg}, S.~{Nobili}, P.~E. {Nugent}, R.~{Pain},
  O.~{Perdereau}, S.~{Perlmutter}, M.~M. {Phillips}, V.~{Prasad}, R.~{Quimby},
  N.~{Regnault}, J.~{Rich}, E.~P. {Rubenstein}, P.~{Ruiz-Lapuente}, F.~D.
  {Santos}, B.~E. {Schaefer}, R.~A. {Schommer}, R.~C. {Smith}, A.~M.
  {Soderberg}, A.~L. {Spadafora}, L.~. {Strolger}, M.~{Strovink}, N.~B.
  {Suntzeff}, N.~{Suzuki}, R.~C. {Thomas}, N.~A. {Walton}, L.~{Wang}, W.~M.
  {Wood-Vasey}, and J.~L. {Yun}, \emph{ArXiv e-prints} \textbf{804} (2008),
  \eprint{0804.4142}.

\bibitem[{Drell} et~al.(2000)]{Drell00}
P.~S. {Drell}, T.~J. {Loredo}, and I.~{Wasserman}, \emph{ApJ} \textbf{530},
  593--617 (2000), \eprint{arXiv:astro-ph/9905027}.

\bibitem[{Aguirre}(1999)]{Aguirre99}
A.~N. {Aguirre}, \emph{ApJ} \textbf{512}, L19--L22 (1999),
  \eprint{arXiv:astro-ph/9811316}.

\bibitem[{Knop} et~al.(2003)]{Knop03}
R.~A. {Knop}, G.~{Aldering}, R.~{Amanullah}, P.~{Astier}, G.~{Blanc}, M.~S.
  {Burns}, A.~{Conley}, S.~E. {Deustua}, M.~{Doi}, R.~{Ellis}, S.~{Fabbro},
  G.~{Folatelli}, A.~S. {Fruchter}, G.~{Garavini}, S.~{Garmond}, K.~{Garton},
  R.~{Gibbons}, G.~{Goldhaber}, A.~{Goobar}, D.~E. {Groom}, D.~{Hardin},
  I.~{Hook}, D.~A. {Howell}, A.~G. {Kim}, B.~C. {Lee}, C.~{Lidman},
  J.~{Mendez}, S.~{Nobili}, P.~E. {Nugent}, R.~{Pain}, N.~{Panagia}, C.~R.
  {Pennypacker}, S.~{Perlmutter}, R.~{Quimby}, J.~{Raux}, N.~{Regnault},
  P.~{Ruiz-Lapuente}, G.~{Sainton}, B.~{Schaefer}, K.~{Schahmaneche},
  E.~{Smith}, A.~L. {Spadafora}, V.~{Stanishev}, M.~{Sullivan}, N.~A. {Walton},
  L.~{Wang}, W.~M. {Wood-Vasey}, and N.~{Yasuda}, \emph{ApJ} \textbf{598},
  102--137 (2003), \eprint{arXiv:astro-ph/0309368}.

\bibitem[{Riess} et~al.(2001)]{Riess_01}
A.~G. {Riess}, P.~E. {Nugent}, R.~L. {Gilliland}, B.~P. {Schmidt}, J.~{Tonry},
  M.~{Dickinson}, R.~I. {Thompson}, T.~{Budav{\'a}ri}, S.~{Casertano}, A.~S.
  {Evans}, A.~V. {Filippenko}, M.~{Livio}, D.~B. {Sanders}, A.~E. {Shapley},
  H.~{Spinrad}, C.~C. {Steidel}, D.~{Stern}, J.~{Surace}, and S.~{Veilleux},
  \emph{ApJ} \textbf{560}, 49--71 (2001), \eprint{arXiv:astro-ph/0104455}.

\bibitem[{Riess} et~al.(2004)]{Riess_04}
A.~G. {Riess}, L.-G. {Strolger}, J.~{Tonry}, S.~{Casertano}, H.~C. {Ferguson},
  B.~{Mobasher}, P.~{Challis}, A.~V. {Filippenko}, S.~{Jha}, W.~{Li},
  R.~{Chornock}, R.~P. {Kirshner}, B.~{Leibundgut}, M.~{Dickinson}, M.~{Livio},
  M.~{Giavalisco}, C.~C. {Steidel}, T.~{Ben{\'{\i}}tez}, and Z.~{Tsvetanov},
  \emph{ApJ} \textbf{607}, 665--687 (2004), \eprint{arXiv:astro-ph/0402512}.

\bibitem[{Riess} et~al.(2007)]{Riess_06}
A.~G. {Riess}, L.-G. {Strolger}, S.~{Casertano}, H.~C. {Ferguson},
  B.~{Mobasher}, B.~{Gold}, P.~J. {Challis}, A.~V. {Filippenko}, S.~{Jha},
  W.~{Li}, J.~{Tonry}, R.~{Foley}, R.~P. {Kirshner}, M.~{Dickinson},
  E.~{MacDonald}, D.~{Eisenstein}, M.~{Livio}, J.~{Younger}, C.~{Xu},
  T.~{Dahl{\'e}n}, and D.~{Stern}, \emph{ApJ} \textbf{659}, 98--121 (2007),
  \eprint{arXiv:astro-ph/0611572}.

\bibitem[{Astier} et~al.(2006)]{SNLS}
P.~{Astier}, J.~{Guy}, N.~{Regnault}, R.~{Pain}, E.~{Aubourg}, D.~{Balam},
  S.~{Basa}, R.~G. {Carlberg}, S.~{Fabbro}, D.~{Fouchez}, I.~M. {Hook}, D.~A.
  {Howell}, H.~{Lafoux}, J.~D. {Neill}, N.~{Palanque-Delabrouille},
  K.~{Perrett}, C.~J. {Pritchet}, J.~{Rich}, M.~{Sullivan}, R.~{Taillet},
  G.~{Aldering}, P.~{Antilogus}, V.~{Arsenijevic}, C.~{Balland}, S.~{Baumont},
  J.~{Bronder}, H.~{Courtois}, R.~S. {Ellis}, M.~{Filiol}, A.~C. {Gon{\c
  c}alves}, A.~{Goobar}, D.~{Guide}, D.~{Hardin}, V.~{Lusset}, C.~{Lidman},
  R.~{McMahon}, M.~{Mouchet}, A.~{Mourao}, S.~{Perlmutter}, P.~{Ripoche},
  C.~{Tao}, and N.~{Walton}, \emph{A \& A} \textbf{447}, 31--48 (2006),
  \eprint{arXiv:astro-ph/0510447}.

\bibitem[{Miknaitis} et~al.(2007)]{Miknaitis_07}
G.~{Miknaitis}, G.~{Pignata}, A.~{Rest}, W.~M. {Wood-Vasey}, S.~{Blondin},
  P.~{Challis}, R.~C. {Smith}, C.~W. {Stubbs}, N.~B. {Suntzeff}, R.~J. {Foley},
  T.~{Matheson}, J.~L. {Tonry}, C.~{Aguilera}, J.~W. {Blackman}, A.~C.
  {Becker}, A.~{Clocchiatti}, R.~{Covarrubias}, T.~M. {Davis}, A.~V.
  {Filippenko}, A.~{Garg}, P.~M. {Garnavich}, M.~{Hicken}, S.~{Jha},
  K.~{Krisciunas}, R.~P. {Kirshner}, B.~{Leibundgut}, W.~{Li}, A.~{Miceli},
  G.~{Narayan}, J.~L. {Prieto}, A.~G. {Riess}, M.~E. {Salvo}, B.~P. {Schmidt},
  J.~{Sollerman}, J.~{Spyromilio}, and A.~{Zenteno}, \emph{ApJ} \textbf{666},
  674--693 (2007), \eprint{arXiv:astro-ph/0701043}.

\bibitem[{Wood-Vasey} et~al.(2007)]{WoodVasey_07}
W.~M. {Wood-Vasey}, G.~{Miknaitis}, C.~W. {Stubbs}, S.~{Jha}, A.~G. {Riess},
  P.~M. {Garnavich}, R.~P. {Kirshner}, C.~{Aguilera}, A.~C. {Becker}, J.~W.
  {Blackman}, S.~{Blondin}, P.~{Challis}, A.~{Clocchiatti}, A.~{Conley},
  R.~{Covarrubias}, T.~M. {Davis}, A.~V. {Filippenko}, R.~J. {Foley},
  A.~{Garg}, M.~{Hicken}, K.~{Krisciunas}, B.~{Leibundgut}, W.~{Li},
  T.~{Matheson}, A.~{Miceli}, G.~{Narayan}, G.~{Pignata}, J.~L. {Prieto},
  A.~{Rest}, M.~E. {Salvo}, B.~P. {Schmidt}, R.~C. {Smith}, J.~{Sollerman},
  J.~{Spyromilio}, J.~L. {Tonry}, N.~B. {Suntzeff}, and A.~{Zenteno},
  \emph{ApJ} \textbf{666}, 694--715 (2007), \eprint{arXiv:astro-ph/0701041}.

\bibitem[{Jha} et~al.(2007)]{JRK_07}
S.~{Jha}, A.~G. {Riess}, and R.~P. {Kirshner}, \emph{ApJ} \textbf{659},
  122--148 (2007), \eprint{arXiv:astro-ph/0612666}.

\bibitem[{Hui} and {Greene}(2006)]{hui_06}
L.~{Hui}, and P.~B. {Greene}, \emph{Phys. Rev. D} \textbf{73}, 123526 (2006),
  \eprint{arXiv:astro-ph/0512159}.

\bibitem[{Frieman} et~al.(2008{\natexlab{b}})]{frieman08}
J.~A. {Frieman}, B.~{Bassett}, A.~{Becker}, C.~{Choi}, D.~{Cinabro},
  F.~{DeJongh}, D.~L. {Depoy}, B.~{Dilday}, M.~{Doi}, P.~M. {Garnavich}, C.~J.
  {Hogan}, J.~{Holtzman}, M.~{Im}, S.~{Jha}, R.~{Kessler}, K.~{Konishi},
  H.~{Lampeitl}, J.~{Marriner}, J.~L. {Marshall}, D.~{McGinnis},
  G.~{Miknaitis}, R.~C. {Nichol}, J.~L. {Prieto}, A.~G. {Riess}, M.~W.
  {Richmond}, R.~{Romani}, M.~{Sako}, D.~P. {Schneider}, M.~{Smith},
  N.~{Takanashi}, K.~{Tokita}, K.~{van der Heyden}, N.~{Yasuda}, C.~{Zheng},
  J.~{Adelman-McCarthy}, J.~{Annis}, R.~J. {Assef}, J.~{Barentine},
  R.~{Bender}, R.~D. {Blandford}, W.~N. {Boroski}, M.~{Bremer},
  H.~{Brewington}, C.~A. {Collins}, A.~{Crotts}, J.~{Dembicky}, J.~{Eastman},
  A.~{Edge}, E.~{Edmondson}, E.~{Elson}, M.~E. {Eyler}, A.~V. {Filippenko},
  R.~J. {Foley}, S.~{Frank}, A.~{Goobar}, T.~{Gueth}, J.~E. {Gunn},
  M.~{Harvanek}, U.~{Hopp}, Y.~{Ihara}, {\v Z}.~{Ivezi{\'c}}, S.~{Kahn},
  J.~{Kaplan}, S.~{Kent}, W.~{Ketzeback}, S.~J. {Kleinman}, W.~{Kollatschny},
  R.~G. {Kron}, J.~{Krzesi{\'n}ski}, D.~{Lamenti}, G.~{Leloudas}, H.~{Lin},
  D.~C. {Long}, J.~{Lucey}, R.~H. {Lupton}, E.~{Malanushenko},
  V.~{Malanushenko}, R.~J. {McMillan}, J.~{Mendez}, C.~W. {Morgan},
  T.~{Morokuma}, A.~{Nitta}, L.~{Ostman}, K.~{Pan}, C.~M. {Rockosi}, A.~K.
  {Romer}, P.~{Ruiz-Lapuente}, G.~{Saurage}, K.~{Schlesinger}, S.~A. {Snedden},
  J.~{Sollerman}, C.~{Stoughton}, M.~{Stritzinger}, M.~{Subba Rao},
  D.~{Tucker}, P.~{Vaisanen}, L.~C. {Watson}, S.~{Watters}, J.~C. {Wheeler},
  B.~{Yanny}, and D.~{York}, \emph{AJ} \textbf{135}, 338--347
  (2008{\natexlab{b}}), \eprint{arXiv:0708.2749}.

\bibitem[{Sako} et~al.(2008)]{sako08}
M.~{Sako}, B.~{Bassett}, A.~{Becker}, D.~{Cinabro}, F.~{DeJongh}, D.~L.
  {Depoy}, B.~{Dilday}, M.~{Doi}, J.~A. {Frieman}, P.~M. {Garnavich}, C.~J.
  {Hogan}, J.~{Holtzman}, S.~{Jha}, R.~{Kessler}, K.~{Konishi}, H.~{Lampeitl},
  J.~{Marriner}, G.~{Miknaitis}, R.~C. {Nichol}, J.~L. {Prieto}, A.~G. {Riess},
  M.~W. {Richmond}, R.~{Romani}, D.~P. {Schneider}, M.~{Smith}, M.~{Subba Rao},
  N.~{Takanashi}, K.~{Tokita}, K.~{van der Heyden}, N.~{Yasuda}, C.~{Zheng},
  J.~{Barentine}, H.~{Brewington}, C.~{Choi}, J.~{Dembicky}, M.~{Harnavek},
  Y.~{Ihara}, M.~{Im}, W.~{Ketzeback}, S.~J. {Kleinman}, J.~{Krzesi{\'n}ski},
  D.~C. {Long}, E.~{Malanushenko}, V.~{Malanushenko}, R.~J. {McMillan},
  T.~{Morokuma}, A.~{Nitta}, K.~{Pan}, G.~{Saurage}, and S.~A. {Snedden},
  \emph{AJ} \textbf{135}, 348--373 (2008), \eprint{arXiv:0708.2750}.

\bibitem[{Guy} et~al.(2005)]{Guy1}
J.~{Guy}, P.~{Astier}, S.~{Nobili}, N.~{Regnault}, and R.~{Pain}, \emph{A \& A}
  \textbf{443}, 781--791 (2005), \eprint{arXiv:astro-ph/0506583}.

\bibitem[{Guy} et~al.(2007)]{Guy2}
J.~{Guy}, P.~{Astier}, S.~{Baumont}, D.~{Hardin}, R.~{Pain}, N.~{Regnault},
  S.~{Basa}, R.~G. {Carlberg}, A.~{Conley}, S.~{Fabbro}, D.~{Fouchez}, I.~M.
  {Hook}, D.~A. {Howell}, K.~{Perrett}, C.~J. {Pritchet}, J.~{Rich},
  M.~{Sullivan}, P.~{Antilogus}, E.~{Aubourg}, G.~{Bazin}, J.~{Bronder},
  M.~{Filiol}, N.~{Palanque-Delabrouille}, P.~{Ripoche}, and
  V.~{Ruhlmann-Kleider}, \emph{A \& A} \textbf{466}, 11--21 (2007),
  \eprint{arXiv:astro-ph/0701828}.

\bibitem[{Frieman} et~al.(2003)]{frieman_03}
J.~A. {Frieman}, D.~{Huterer}, E.~V. {Linder}, and M.~S. {Turner}, \emph{Phys.
  Rev. D} \textbf{67}, 083505 (2003), \eprint{arXiv:astro-ph/0208100}.

\bibitem[Boughn and Crittenden(2004)]{Boughn_Crittenden}
S.~Boughn, and R.~Crittenden, \emph{Nature} \textbf{427}, 45--47 (2004),
  \eprint{astro-ph/0305001}.

\bibitem[Fosalba and Gaztanaga(2004)]{Fosalba_Gaztanaga}
P.~Fosalba, and E.~Gaztanaga, \emph{Mon. Not. Roy. Astron. Soc.} \textbf{350},
  L37--L41 (2004), \eprint{astro-ph/0305468}.

\bibitem[Afshordi et~al.(2004)]{Afshordi_Strauss}
N.~Afshordi, Y.-S. Loh, and M.~A. Strauss, \emph{Phys. Rev.} \textbf{D69},
  083524 (2004), \eprint{astro-ph/0308260}.

\bibitem[{Scranton} et~al.(2003)]{Scranton_ISW}
R.~{Scranton}, A.~J. {Connolly}, R.~C. {Nichol}, A.~{Stebbins}, I.~{Szapudi},
  D.~J. {Eisenstein}, N.~{Afshordi}, T.~{Budavari}, I.~{Csabai}, J.~A.
  {Frieman}, J.~E. {Gunn}, D.~{Johnson}, Y.~{Loh}, R.~H. {Lupton}, C.~J.
  {Miller}, E.~S. {Sheldon}, R.~S. {Sheth}, A.~S. {Szalay}, M.~{Tegmark}, and
  Y.~{Xu}, \emph{astro-ph/0307335}  (2003), \eprint{astro-ph/0307335}.

\bibitem[{Schneider}(2006)]{Schneider_review}
P.~{Schneider}, \enquote{{Part 3: Weak gravitational lensing},} in
  \emph{Saas-Fee Advanced Course 33: Gravitational Lensing: Strong, Weak and
  Micro}, edited by G.~{Meylan}, P.~{Jetzer}, P.~{North}, P.~{Schneider}, C.~S.
  {Kochanek}, and J.~{Wambsganss}, 2006, pp. 269--451.

\bibitem[Munshi et~al.(2006)]{Munshi_review}
D.~Munshi, P.~Valageas, L.~Van~Waerbeke, and A.~Heavens,
  \emph{astro-ph/0612667}  (2006), \eprint{astro-ph/0612667}.

\bibitem[{Bacon} et~al.(2000)]{bacon00}
D.~J. {Bacon}, A.~R. {Refregier}, and R.~S. {Ellis}, \emph{MNRAS} \textbf{318},
  625--640 (2000), \eprint{arXiv:astro-ph/0003008}.

\bibitem[{Kaiser} et~al.(2000)]{kaiser00}
N.~{Kaiser}, G.~{Wilson}, and G.~A. {Luppino}, \emph{astro-ph/0003338}  (2000),
  \eprint{astro-ph/0003338}.

\bibitem[{Van Waerbeke} et~al.(2000)]{vanW00}
L.~{Van Waerbeke}, Y.~{Mellier}, T.~{Erben}, J.~C. {Cuillandre},
  F.~{Bernardeau}, R.~{Maoli}, E.~{Bertin}, H.~J. {Mc Cracken}, O.~{Le
  F{\`e}vre}, B.~{Fort}, M.~{Dantel-Fort}, B.~{Jain}, and P.~{Schneider},
  \emph{A \& A} \textbf{358}, 30--44 (2000), \eprint{arXiv:astro-ph/0002500}.

\bibitem[{Wittman} et~al.(2000)]{Witt00}
D.~M. {Wittman}, J.~A. {Tyson}, D.~{Kirkman}, I.~{Dell'Antonio}, and
  G.~{Bernstein}, \emph{Nature} \textbf{405}, 143--148 (2000),
  \eprint{arXiv:astro-ph/0003014}.

\bibitem[Jarvis et~al.(2006)]{Jarvis_CTIO}
M.~Jarvis, B.~Jain, G.~Bernstein, and D.~Dolney, \emph{ApJ} \textbf{644},
  71--79 (2006), \eprint{astro-ph/0502243}.

\bibitem[{Hoekstra} et~al.(2006)]{Hoekstra}
H.~{Hoekstra}, Y.~{Mellier}, L.~{van Waerbeke}, E.~{Semboloni}, L.~{Fu}, M.~J.
  {Hudson}, L.~C. {Parker}, I.~{Tereno}, and K.~{Benabed}, \emph{ApJ}
  \textbf{647}, 116--127 (2006), \eprint{arXiv:astro-ph/0511089}.

\bibitem[{Massey} et~al.(2007)]{Massey}
R.~{Massey}, J.~{Rhodes}, R.~{Ellis}, N.~{Scoville}, A.~{Leauthaud},
  A.~{Finoguenov}, P.~{Capak}, D.~{Bacon}, H.~{Aussel}, J.-P. {Kneib},
  A.~{Koekemoer}, H.~{McCracken}, B.~{Mobasher}, S.~{Pires}, A.~{Refregier},
  S.~{Sasaki}, J.-L. {Starck}, Y.~{Taniguchi}, A.~{Taylor}, and J.~{Taylor},
  \emph{Nature} \textbf{445}, 286--290 (2007), \eprint{arXiv:astro-ph/0701594}.

\bibitem[Huterer(2002)]{Huterer_thesis}
D.~Huterer, \emph{Phys. Rev.} \textbf{D65}, 063001 (2002),
  \eprint{astro-ph/0106399}.

\bibitem[{Hu}(2002)]{Hu02}
W.~{Hu}, \emph{Phys. Rev. D} \textbf{65}, 023003 (2002),
  \eprint{arXiv:astro-ph/0108090}.

\bibitem[Allen et~al.(2004)]{Allen_04}
S.~W. Allen, R.~W. Schmidt, H.~Ebeling, A.~C. Fabian, and L.~van Speybroeck,
  \emph{Mon. Not. Roy. Astron. Soc.} \textbf{353}, 457 (2004),
  \eprint{astro-ph/0405340}.

\bibitem[{Allen} et~al.(2007)]{Allen_07}
S.~W. {Allen}, D.~A. {Rapetti}, R.~W. {Schmidt}, H.~{Ebeling}, G.~{Morris}, and
  A.~C. {Fabian}, \emph{astro-ph/0706.0033}  (2007), \eprint{0706.0033}.

\bibitem[{Fukugita} et~al.(1992)]{fukugita92}
M.~{Fukugita}, T.~{Futamase}, M.~{Kasai}, and E.~L. {Turner}, \emph{ApJ}
  \textbf{393}, 3--21 (1992).

\bibitem[{Kochanek}(1996)]{Kochanek_96}
C.~S. {Kochanek}, \emph{ApJ} \textbf{466}, 638 (1996),
  \eprint{arXiv:astro-ph/9510077}.

\bibitem[{Mitchell} et~al.(2005)]{mitchell05}
J.~L. {Mitchell}, C.~R. {Keeton}, J.~A. {Frieman}, and R.~K. {Sheth},
  \emph{ApJ} \textbf{622}, 81--98 (2005), \eprint{arXiv:astro-ph/0401138}.

\bibitem[{Oguri} et~al.(2007)]{oguri07}
M.~{Oguri}, N.~{Inada}, M.~A. {Strauss}, C.~S. {Kochanek}, G.~T. {Richards},
  D.~P. {Schneider}, R.~H. {Becker}, M.~{Fukugita}, M.~D. {Gregg}, P.~B.
  {Hall}, J.~F. {Hennawi}, D.~E. {Johnston}, I.~{Kayo}, C.~R. {Keeton},
  B.~{Pindor}, M.-S. {Shin}, E.~L. {Turner}, R.~L. {White}, D.~G. {York}, S.~F.
  {Anderson}, N.~A. {Bahcall}, R.~J. {Brunner}, S.~{Burles}, F.~J. {Castander},
  K.~{Chiu}, A.~{Clocchiatti}, D.~{Eisenstein}, J.~A. {Frieman}, Y.~{Kawano},
  R.~{Lupton}, T.~{Morokuma}, H.-W. {Rix}, R.~{Scranton}, and E.~S. {Sheldon},
  \emph{ArXiv:astro-ph/0708.0825}  (2007), \eprint{0708.0825}.

\bibitem[{Chae}(2007)]{chae07}
K.-H. {Chae}, \emph{ApJ} \textbf{658}, L71--L74 (2007),
  \eprint{arXiv:astro-ph/0611898}.

\bibitem[Weinberg(1987)]{Weinberg87}
S.~Weinberg, \emph{Phys. Rev. Lett.} \textbf{59}, 2607 (1987).

\bibitem[Bousso and Polchinski(2000)]{Bousso_Polchinski}
R.~Bousso, and J.~Polchinski, \emph{JHEP} \textbf{06}, 006 (2000),
  \eprint{hep-th/0004134}.

\bibitem[Susskind(2003)]{Susskind}
L.~Susskind, \emph{hep-th/0302219}  (2003), \eprint{hep-th/0302219}.

\bibitem[Ratra and Peebles(1988)]{Ratra_Peebles}
B.~Ratra, and P.~J.~E. Peebles, \emph{Phys. Rev.} \textbf{D37}, 3406 (1988).

\bibitem[Wetterich(1988)]{Wetterich}
C.~Wetterich, \emph{Nucl. Phys.} \textbf{B302}, 668 (1988).

\bibitem[Zlatev et~al.(1999)]{Zlatev}
I.~Zlatev, L.-M. Wang, and P.~J. Steinhardt, \emph{Phys. Rev. Lett.}
  \textbf{82}, 896--899 (1999), \eprint{astro-ph/9807002}.

\bibitem[Coble et~al.(1997)]{Coble}
K.~Coble, S.~Dodelson, and J.~A. Frieman, \emph{Phys. Rev.} \textbf{D55},
  1851--1859 (1997), \eprint{astro-ph/9608122}.

\bibitem[{Nomura} et~al.(2000)]{Nomura}
Y.~{Nomura}, T.~{Watari}, and T.~{Yanagida}, \emph{Phys. Rev. D.} \textbf{61},
  105007--+ (2000), \eprint{arXiv:hep-ph/9911324}.

\bibitem[Hall et~al.(2005)]{Hall}
L.~J. Hall, Y.~Nomura, and S.~J. Oliver, \emph{Phys. Rev. Lett.} \textbf{95},
  141302 (2005), \eprint{astro-ph/0503706}.

\bibitem[Caldwell and Linder(2005)]{Caldwell_Linder}
R.~R. Caldwell, and E.~V. Linder, \emph{Phys. Rev. Lett.} \textbf{95}, 141301
  (2005), \eprint{astro-ph/0505494}.

\bibitem[Dvali et~al.(2000)]{DGP}
G.~R. Dvali, G.~Gabadadze, and M.~Porrati, \emph{Phys. Lett.} \textbf{B485},
  208--214 (2000), \eprint{hep-th/0005016}.

\bibitem[Deffayet(2001)]{Deffayet}
C.~Deffayet, \emph{Phys. Lett.} \textbf{B502}, 199--208 (2001),
  \eprint{hep-th/0010186}.

\bibitem[Carroll et~al.(2004)]{CDTT}
S.~M. Carroll, V.~Duvvuri, M.~Trodden, and M.~S. Turner, \emph{Phys. Rev.}
  \textbf{D70}, 043528 (2004), \eprint{astro-ph/0306438}.

\bibitem[Song et~al.(2007)]{Song_Hu_fR}
Y.-S. Song, W.~Hu, and I.~Sawicki, \emph{Phys. Rev.} \textbf{D75}, 044004
  (2007), \eprint{astro-ph/0610532}.

\bibitem[Gregory et~al.(2007)]{Gregory07}
R.~Gregory, N.~Kaloper, R.~C. Myers, and A.~Padilla, \emph{JHEP} \textbf{10},
  069 (2007), \eprint{arXiv:0707.2666 [hep-th]}.

\bibitem[Kolb et~al.(2006)]{Kolb_Matarrese_06}
E.~W. Kolb, S.~Matarrese, and A.~Riotto, \emph{New J. Phys.} \textbf{8}, 322
  (2006), \eprint{astro-ph/0506534}.

\bibitem[Tomita(2001)]{Tomita_00}
K.~Tomita, \emph{Mon. Not. Roy. Astron. Soc.} \textbf{326}, 287 (2001),
  \eprint{astro-ph/0011484}.

\bibitem[Alnes et~al.(2006)]{Alnes_05}
H.~Alnes, M.~Amarzguioui, and O.~Gron, \emph{Phys. Rev.} \textbf{D73}, 083519
  (2006), \eprint{astro-ph/0512006}.

\bibitem[Enqvist(2007)]{Enqvist_07}
K.~Enqvist, \emph{astro-ph/0709.2044}  (2007), \eprint{arXiv:0709.2044
  [astro-ph]}.

\bibitem[{Caldwell} and {Stebbins}(2008)]{CaldSteb}
R.~R. {Caldwell}, and A.~{Stebbins}, \emph{Physical Review Letters}
  \textbf{100}, 191302--+ (2008), \eprint{arXiv:0711.3459}.

\bibitem[{Warren} et~al.(2006)]{warren_06}
M.~S. {Warren}, K.~{Abazajian}, D.~E. {Holz}, and L.~{Teodoro}, \emph{ApJ}
  \textbf{646}, 881--885 (2006), \eprint{arXiv:astro-ph/0506395}.

\bibitem[{Tinker} et~al.(2008)]{tinker08}
J.~L. {Tinker}, A.~V. {Kravtsov}, A.~{Klypin}, K.~{Abazajian}, M.~S. {Warren},
  G.~{Yepes}, S.~{Gottlober}, and D.~E. {Holz}, \emph{ArXiv e-prints}
  \textbf{803} (2008), \eprint{0803.2706}.

\bibitem[{Wang} and {Steinhardt}(1998)]{wang_98}
L.~{Wang}, and P.~J. {Steinhardt}, \emph{ApJ} \textbf{508}, 483--490 (1998),
  \eprint{arXiv:astro-ph/9804015}.

\bibitem[{Haiman} et~al.(2001)]{haiman_01}
Z.~{Haiman}, J.~J. {Mohr}, and G.~P. {Holder}, \emph{ApJ} \textbf{553},
  545--561 (2001), \eprint{arXiv:astro-ph/0002336}.

\bibitem[{Borgani}(2006)]{borgani_06}
S.~{Borgani}, \emph{astro-ph/0605575}  (2006), \eprint{astro-ph/0605575}.

\bibitem[{Mohr}(2005)]{Mohr_04}
J.~J. {Mohr}, \enquote{{Cluster Survey Studies of the Dark Energy},} in
  \emph{Observing Dark Energy}, edited by S.~C. {Wolff}, and T.~R. {Lauer},
  2005, vol. 339 of \emph{Astronomical Society of the Pacific Conference
  Series}, p. 140.

\bibitem[{Majumdar} and {Mohr}(2004)]{Majumdar_04}
S.~{Majumdar}, and J.~J. {Mohr}, \emph{ApJ} \textbf{613}, 41--50 (2004),
  \eprint{arXiv:astro-ph/0305341}.

\bibitem[{Lima} and {Hu}(2004)]{Lima_04}
M.~{Lima}, and W.~{Hu}, \emph{Phys. Rev. D} \textbf{70}, 043504 (2004),
  \eprint{arXiv:astro-ph/0401559}.

\bibitem[{Kaiser}(1992)]{Kaiser92}
N.~{Kaiser}, \emph{ApJ} \textbf{388}, 272--286 (1992).

\bibitem[{Hu} and {Jain}(2004)]{hujain03}
W.~{Hu}, and B.~{Jain}, \emph{Phys. Rev. D} \textbf{70}, 043009 (2004),
  \eprint{arXiv:astro-ph/0312395}.

\bibitem[{Jain} and {Taylor}(2003)]{JainTaylor}
B.~{Jain}, and A.~{Taylor}, \emph{Phys. Rev. Lett.} \textbf{91}, 141302 (2003),
  \eprint{arXiv:astro-ph/0306046}.

\bibitem[Bernstein and Jain(2004)]{Bernstein_Jain}
G.~M. Bernstein, and B.~Jain, \emph{ApJ} \textbf{600}, 17--25 (2004),
  \eprint{astro-ph/0309332}.

\bibitem[Takada and Jain(2004)]{Takada_Jain}
M.~Takada, and B.~Jain, \emph{Mon. Not. Roy. Astron. Soc.} \textbf{348}, 897
  (2004), \eprint{astro-ph/0310125}.

\bibitem[{Huterer} et~al.(2006)]{Huterer06}
D.~{Huterer}, M.~{Takada}, G.~{Bernstein}, and B.~{Jain}, \emph{MNRAS}
  \textbf{366}, 101--114 (2006), \eprint{arXiv:astro-ph/0506030}.

\bibitem[{Ma} et~al.(2006)]{Ma06}
Z.~{Ma}, W.~{Hu}, and D.~{Huterer}, \emph{ApJ} \textbf{636}, 21--29 (2006),
  \eprint{arXiv:astro-ph/0506614}.

\bibitem[{Hirata} and {Seljak}(2004)]{Hirata04}
C.~M. {Hirata}, and U.~{Seljak}, \emph{Phys. Rev. D} \textbf{70}, 063526
  (2004), \eprint{arXiv:astro-ph/0406275}.

\bibitem[{Heymans} et~al.(2006)]{Heymans}
C.~{Heymans}, M.~{White}, A.~{Heavens}, C.~{Vale}, and L.~{van Waerbeke},
  \emph{MNRAS} \textbf{371}, 750--760 (2006), \eprint{arXiv:astro-ph/0604001}.

\bibitem[{Zentner} et~al.(2007)]{zentner07}
A.~R. {Zentner}, D.~H. {Rudd}, and W.~{Hu}, \emph{astro-ph/0709.4029}  (2007),
  \eprint{0709.4029}.

\bibitem[{Albrecht} et~al.(2006)]{DETF}
A.~{Albrecht}, G.~{Bernstein}, R.~{Cahn}, W.~L. {Freedman}, J.~{Hewitt},
  W.~{Hu}, J.~{Huth}, M.~{Kamionkowski}, E.~W. {Kolb}, L.~{Knox}, J.~C.
  {Mather}, S.~{Staggs}, and N.~B. {Suntzeff}, \emph{astro-ph/0609591}  (2006).

\end{thebibliography}

\IfFileExists{\jobname.bbl}{}
 {\typeout{}
  \typeout{******************************************}
  \typeout{** Please run "bibtex \jobname" to optain}
  \typeout{** the bibliography and then re-run LaTeX}
  \typeout{** twice to fix the references!}
  \typeout{******************************************}
  \typeout{}
 }

\end{document}